%% file: degg_draft_v3.3.tex
\documentclass[a4paper,11pt]{article}
\usepackage{jinstpub} 
\usepackage[separate-uncertainty]{siunitx}
\usepackage{isotope}
\DeclareSIUnit{\PE}{PE}
\DeclareSIUnit{\SPS}{SPS}
\DeclareSIUnit{\bitpersecond}{bps}
\usepackage{placeins}
\usepackage{multirow}
\DeclareUnicodeCharacter{2212}{−}
\usepackage{lineno}

\compress 
\title{D-Egg: a Dual PMT Optical Module for IceCube}

\collaboration{The IceCube Collaboration}

\input{authors.tex}
\input{extra_authors.tex}

\emailAdd{analysis@icecube.wisc.edu}

\abstract{
The D-Egg, an acronym for ``Dual optical sensors in an Ellipsoid Glass for Gen2,'' is one of the optical modules designed for future extensions of the IceCube experiment at the South Pole. 
The D-Egg has an elongated-sphere shape to maximize the photon-sensitive effective area while maintaining a narrow diameter to reduce the cost and the time needed for drilling of the deployment holes in the glacial ice for the optical modules at depths up to \SI{2700}{\metre}.

The D-Egg design is utilized for the IceCube Upgrade, the next stage of the IceCube project also known as IceCube-Gen2 Phase 1, where nearly half of the optical sensors to be deployed are D-Eggs. With two 8-inch high-quantum efficiency photomultiplier tubes (PMTs) per module, D-Eggs offer an increased effective area while retaining the successful design of the IceCube digital optical module (DOM). The convolution of the wavelength-dependent effective area and the Cherenkov emission spectrum provides an effective photodetection sensitivity that is \num{2.8}~times larger than that of IceCube DOMs. 
The signal of each of the two PMTs is digitized using ultra-low-power \num{14}-bit {analog-to-digital converters} with a sampling frequency of \SI{240}{\mega\SPS}, enabling a flexible event triggering, as well as seamless and lossless event recording of single-photon signals to multi-photons exceeding \num{200}~photoelectrons within \SI{10}{\nano\second}. 
Mass production of D-Eggs has been completed, with \num{277} out of the \num{310}~\mbox{D-Eggs} produced to be used in the IceCube Upgrade. 
In this paper, we report the design of the D-Eggs, as well as the sensitivity and the single to multi-photon detection performance of mass-produced D-Eggs measured in a laboratory 
using the built-in data acquisition system in each D-Egg optical sensor module.
}


\begin{document}
\maketitle
\flushbottom

\section{IceCube and the IceCube Upgrade}\label{sec:introduction}
The IceCube Neutrino Observatory~\cite{icecube_instrumentation}, located at the geographic South Pole, was completed in 2010 and began full operation in 2011. IceCube has led to many new findings in high-energy astroparticle physics, including the discovery of a {diffuse} astrophysical neutrino flux~\cite{Aartsen:2013bka,Aartsen:2013jdh}, temporal and directional correlation of neutrinos with a flaring blazar~\cite{IceCube:2018dnn}, steady emission from an active galactic nucleus~\cite{IceCube:2019cia, IceCube:2022sci}, TeV neutrino interaction cross-section~\cite{IceCube:2017mmn}, and a Glashow resonance event candidate~\cite{IceCube:2021rpz}.
Furthermore, IceCube collects atmospheric neutrino data at energies as low as $\sim\SI{5}{\giga \electronvolt}$ to measure atmospheric muon neutrino disappearance~\cite{Aartsen:2018osc} and tau neutrino appearance~\cite{Aartsen:2019osctau} to extract neutrino oscillation parameters with high statistics. 

IceCube uses highly transparent glacial ice as a Cherenkov medium for detecting secondary charged particles produced in neutrino interactions with the Earth. Energy, direction, and flavor of incoming neutrinos are reconstructed from the timing and spatial distribution of Cherenkov light measured with a \SI{1}{\kilo\metre\cubed} array of \num{5160}~optical sensors deployed on \num{86}~cables, called strings, installed at depths between \SIlist{1450;2450}{\metre}. Optical sensors for IceCube are called digital optical modules (DOMs), the primary component of which is a downward-facing photomultiplier tube (PMT). Eight of the \num{86}~strings belong to a densely instrumented sub-array called DeepCore~\cite{deepcore}. 
While the South Pole is considered one of the world's harshest environments, glacial ice $\sim\SI{2}{\kilo\metre}$ below the surface is a dark and transparent environment with stable temperature profiles ideal for noise-sensitive optical sensors. 
The average detector uptime of IceCube in the last several years is higher than \SI{99.8}{\percent}.
Although it has been \num{15}~years since the first installation of the sensors, the optical modules have been found to have an extremely low failure rate ($<\SI{1.7}{\percent}$ over the first \num{15}~years of the operation)~\cite{icecube_instrumentation}, demonstrating that the transparent, cold, dark, and environmentally stable South Pole ice is a suitable location for a Cherenkov neutrino telescope. 

As extensions to the current IceCube detector, two projects are underway: the IceCube Upgrade~\cite{Ishihara:2019aao}, which will significantly increase IceCube's sensitivity at low energies, 
and IceCube-Gen2, which is planned to be a high-energy extension of IceCube~\cite{IceCube-Gen2:2020qha}. 
The IceCube Upgrade consists of \num{7}~new strings instrumented with approximately \num{700}~optical sensors in a grid spacing of \SI{20}{\metre}~(horizontal)~$\times~\SI{3}{\metre}$~(vertical). The sensors are deployed at depths between \SIlist{2150;2425}{\metre}, where the glacial ice is the clearest, and the atmospheric muon background is reduced. 
A few optical sensors will also be installed in shallower and deeper regions. \mbox{IceCube-Gen2} is a planned large-scale high-energy extension of IceCube optimized to search for sources of astrophysical neutrinos from \si{\tera\electronvolt} to \si{\exa\electronvolt} energies. 
%
%
%
The baseline design of the IceCube-Gen2 optical array instruments an \SI{8}{\kilo\metre\cubed} volume with a grid spacing  of \SI{240}{\metre}~(horizontal)~$\times~\SI{17}{\metre}$~(vertical). It consists of \num{120}~holes and considers drilling with a smaller diameter than in IceCube as a feasible option for saving fuel costs and drill time~\cite{IceCube-Gen2:2020qha}.
%
%
This \mbox{IceCube-Gen2} design, therefore, sets new requirements for optical sensors compared to previous projects, 
i.e.\ to make the module fit into the reduced diameter while maintaining high sensitivity.

The primary optical modules to be installed in the IceCube Upgrade array are the multi-PMT digital optical module~(mDOM)~\cite{Classen:2017gz,vanEijk:2019G1,anderson2021design} and the D-Egg, an acronym for ``Dual optical sensors in an Ellipsoid Glass for Gen2.''
Both mDOMs and D-Eggs are distributed throughout the IceCube Upgrade array. 
Their locations have been selected for the best sensitivity for \si{\giga\electronvolt}-energy neutrinos and calibration of the optical properties of glacial ice. Also, the shallower and deeper regions of the IceCube upgrade array are suitable as test sites of the future project, IceCube-Gen2~\cite{IceCube-Gen2:2020qha}. A total of \num{310}~\mbox{D-Eggs} {have} been assembled by the end of 2021, of which \num{277} are planned to be deployed during the 2025/2026 South Pole season~\cite{Ishihara:2019aao}. After mass production, D-Eggs entered mass testing prior to shipping to the South Pole. 
%
In this paper, we describe the design of the D-Egg as well as the performance of these mass-produced D-Eggs as measured in a laboratory freezer setup. 
The paper is organized as follows. Section~\ref{sec:degg} describes the structure and components of the D-Egg, and Section~\ref{sec:eff_area} presents its sensitivity. In Section~\ref{sec:degg_performance}, we describe the results of the acceptance testing in the laboratory. Finally, Section~\ref{sec:conclusion} concludes the paper.

\section{D-Egg Optical Modules}\label{sec:degg}
The D-Egg has been designed and optimized to improve the sensitivity and uniformity in detecting Cherenkov photons compared to the IceCube DOMs while allowing for a reduction of the diameter of the deployment holes~\cite{Shimizu:2017rcd}. The diameter reduction is motivated by simulations and past experiences of large-scale drilling, which indicate that a diameter reduction of \SI{5}{\centi\metre} would allow savings of \SI{15}{\percent} in drill time and fuel consumption~\cite{IceCube-Gen2:2020qha}. 

\begin{figure}[t]
\centering
\includegraphics[height=0.26\textheight,trim={{1.4\textwidth} 0 {1.4\textwidth} 0},clip]{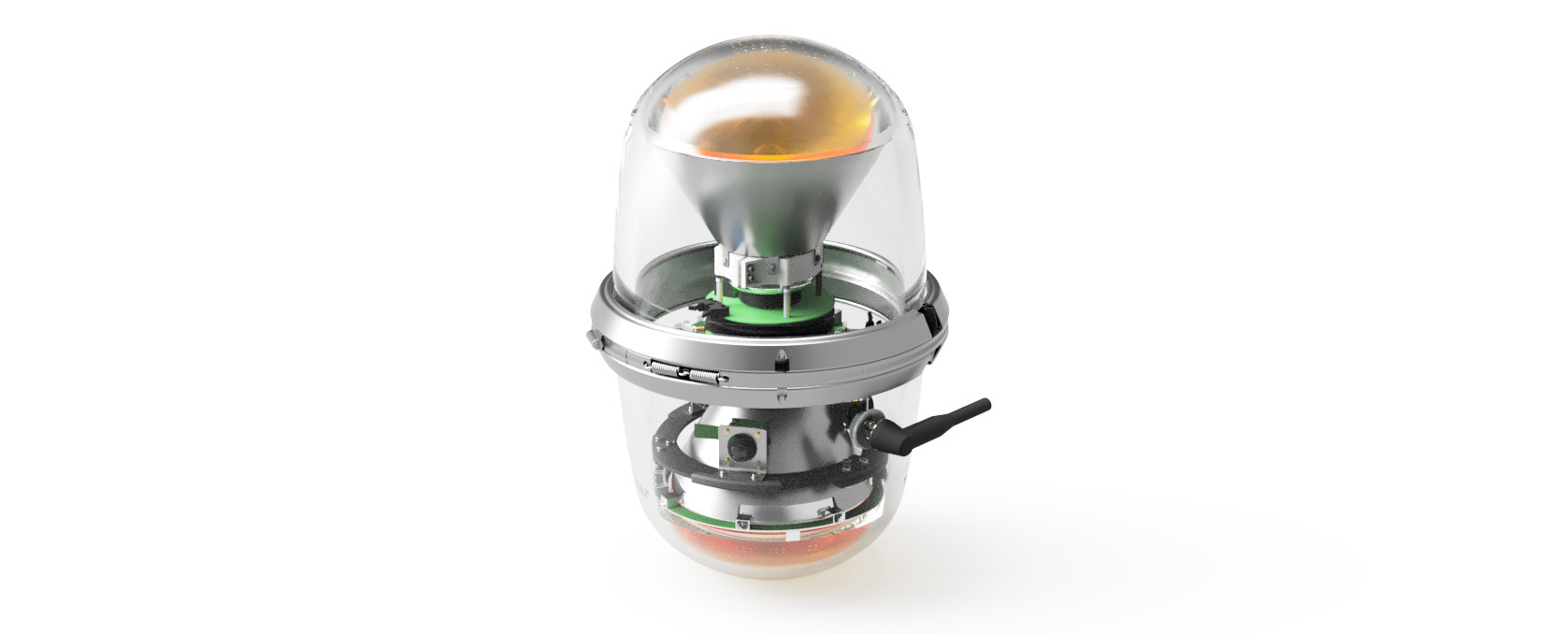}~
\includegraphics[height=0.26\textheight,clip]{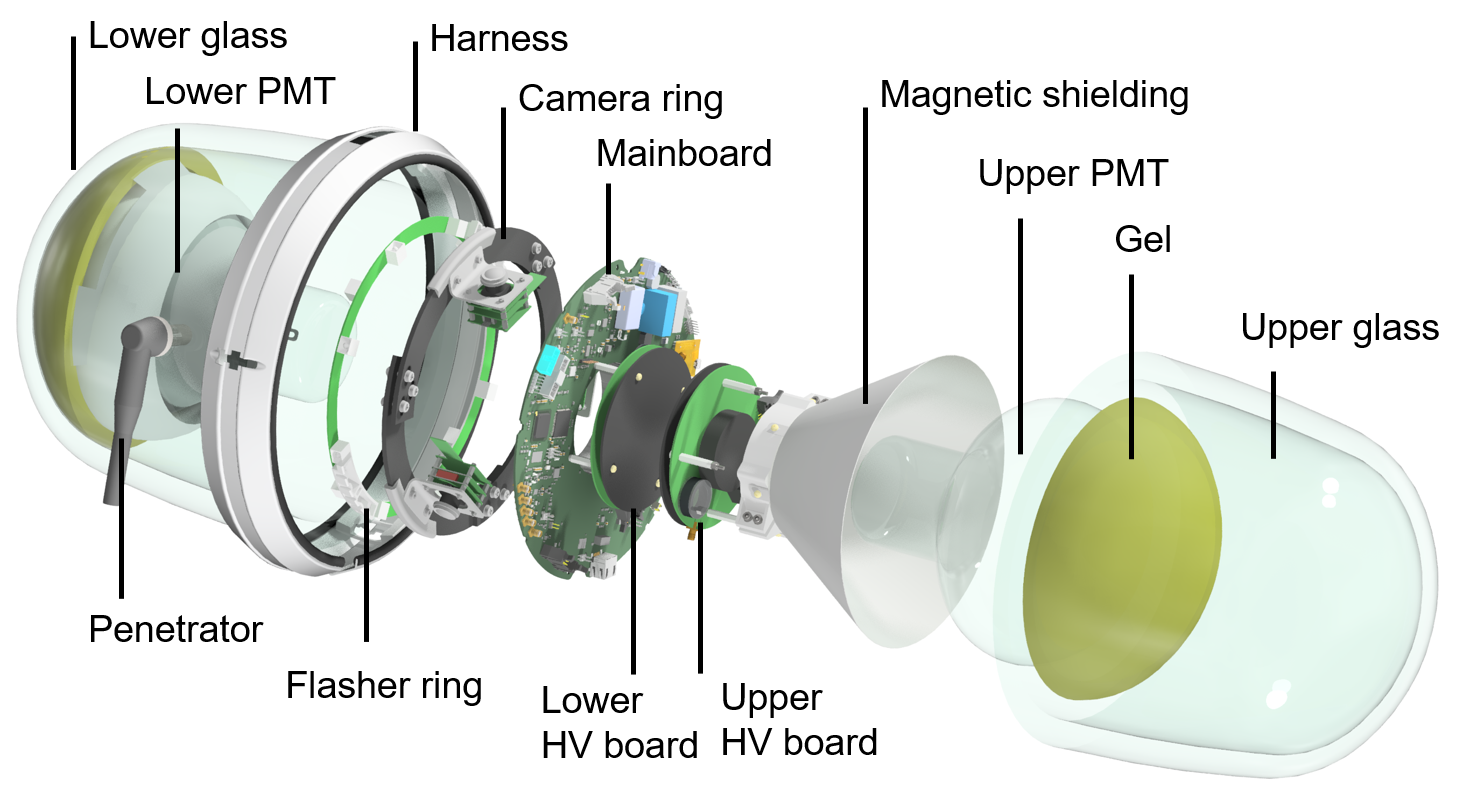}
\caption{Images of the D-Egg module designed and built for the IceCube Upgrade. 
The left figure shows a \mbox{D-Egg} with its sealed UV-transparent glass housing and harness around the equator, which is used to hold the \mbox{D-Egg} during deployment. 
The right figure is an exploded view showing the \mbox{D-Egg} internal structure, including the mainboard, calibration devices (cameras and LED flashers), PMTs, optical coupling silicone (gel), and magnetic shielding (FINEMET$^{\text{\textregistered}}$).}
\label{fig:tmp_degg}
\end{figure}

The left panel of Figure~\ref{fig:tmp_degg} presents a 3D CAD image of a fully assembled D-Egg, with the component structure shown in the right panel.
Two 8-inch high-quantum-efficiency PMTs from Hamamatsu Photonics K.K.~(HPK)\footnote{\url{http://www.hamamatsu.com}} are enclosed in an elongated-sphere-shaped, UV-transparent, and pressure-resistant glass housing.
The upward and downward-facing PMTs are supported and optically coupled to the glass housing with a silicone elastomer. The necks of the two PMTs are connected to each other by a silicone buffer sheet sandwiched between two high-voltage (HV) boards. 
Calibration devices, such as calibrated light sources~(LED flashers)~\cite{AARTSEN201373,Ayumi-ICRC} and optical cameras~\cite{IceCube:2021jfx}, 
are installed in an open space in the lower hemisphere. 
The D-Egg glass housing is sealed during assembly and is held using an external aluminum belt~(harness) surrounding the center of the pressure glass and three steel ropes attached to the harness.

Each of the two PMTs is attached to a PMT HV board consisting of an HV generator, a voltage divider circuit, and a pulse-shaping circuit. The readout, control, and communication circuits are located on a central electrical board called the mainboard. The disk-shaped mainboard is placed in the bottom hemisphere near the equator. 
%
A single wire pair establishes power and bidirectional digital communication between the mainboard and the computers in the IceCube Laboratory~(ICL)~\cite{Kelley_2021}. Two other wire pairs are connected to the mainboard to indicate the D-Egg address. These three pairs of wires are fed to the outside through a \SI{16.5}{\milli\metre} bushing in the pressure vessel with a so-called penetrator cable. The penetrator cable is connected to a breakout cable that comes off the main vertical cable. The main cable connects to a communication and power hub, called FieldHub, in the ICL.
%

\subsection{D-Egg Photomultiplier Tube}

Two 8-inch \mbox{R5912-100-70} high-quantum-efficiency PMTs from HPK are enclosed in the pressure vessel. The photon detection uniformity of the D-Egg with upper and lower PMTs is significantly improved compared to the IceCube DOM, which has a single 10-inch \mbox{R7081-02} PMT from HPK\@. 
In addition to its physical dimensions, this PMT was selected based on criteria of low dark noise, good time and charge resolution for single photoelectron (SPE) signals, and high quantum efficiency~(QE), suitable for detecting Cherenkov photons. The \mbox{R5912-100-70} has ten linearly focused dynode stages and achieves a nominal gain of \num{e7} at high voltage values typically between \SIlist{1300;1800}{\volt}. Each PMT has a specific high-voltage setting for a given temperature to achieve a gain of \num{e7}. 

The \mbox{R5912-100-70} PMT features an improved QE compared to that of the 10-inch PMT \mbox{R7081-02}~\cite{IceCube:2010dpc}, with a photocathode composed of super bialkali material sensitive to photons with wavelengths between \SIlist{300;650}{\nano\metre} with a typical peak QE of approximately \SI{36}{\percent} at \SI{350}{\nano\metre}.
The typical dark rate of the PMTs at low temperatures is approximately \SI{450}{\hertz}, which is $\sim\num{1.5}$~times higher than that of the \mbox{R7081-02}. This factor is mostly attributed to the higher QE of the \mbox{R5912-100-70}, which reaches approximately \SI{25}{\percent} at wavelengths between \SIlist{350;400}{\nano\metre}.

\subsection{Pressure Vessel and Optical Coupling Silicone}
As shown in the left panel of Figure~\ref{fig:tmp_degg}, the D-Egg utilizes an elongated-sphere-shaped glass housing. Each half of the pressure vessel consists of a hemisphere whose internal radius matches the surface curvature of the \mbox{R5912-100-70} PMTs, taking into account the additional \SI{5}{\milli\metre} {for} {the} optical coupling silicone, and a cylindrical-like structure with a diameter and thickness gradually increasing towards the equator. The glass housing has an outer diameter of \SI{300}{\milli\metre} and a height of \SI{534}{\milli\metre}. The glass thickness is \SI{10}{\milli\metre} at the top and bottom and \SI{20}{\milli\metre} at the equator. The size of the housing is optimized to fit both the upper and lower PMTs with the HV boards back-to-back. 
The thickness of the glass in front of the PMT photocathodes is minimized for better sensitivity, and thicker glass is utilized on the sides for {greater} mechanical strength. 

The pressure vessel is required to withstand a maximum pressure of \SI{70}{\mega\pascal} during the refreezing of a drilled hole~\cite{icecube_instrumentation}. A series of pressure tests up to \SI{72}{\mega\pascal} were performed on prototype D-Egg glass housings with and without enclosed PMTs in a hyperbaric chamber at JAMSTEC.\footnote{Japan Agency for Marine-Earth Science and Technology; \url{http://www.jamstec.go.jp/e}} The mechanical compression of the glass under high pressure was measured by using three strain gauges attached to the inside of the glass housing. The measured strain was proportional to the applied pressure following Hooke's law and in good agreement with the expected strain from simulations. The horizontal diameter of the glass vessel at the equator is reduced by approximately \SI{2}{\milli\metre} under \SI{70}{\mega\pascal}.

\begin{figure}[t]
\centering
\includegraphics[width=0.7\textwidth]{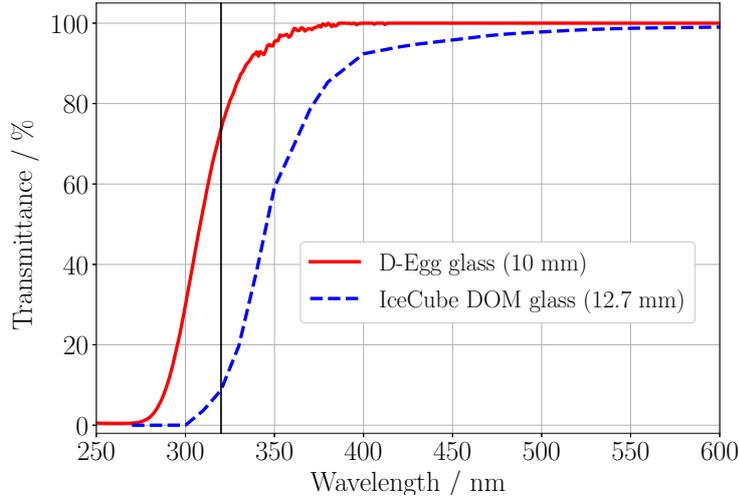}
\caption{Light transmittance as a function of wavelength for glasses in the 
D-Egg and the IceCube~DOM. Due to the careful selection of UV-transparent components, the transmittance is greatly enhanced at shorter wavelengths. 
The wavelength of \SI{320}{\nano\metre} (the manufacturer's specification wavelength) is shown as a vertical line.
}
\label{glass_properties}
\end{figure}

In addition to mechanical strength, the glass vessel is designed such that the D-Egg satisfies the scientific requirements of uniform solid angle sensitivity, high photon detection efficiency, and low dark rate.
Uniformity is improved by matching the glass shape to the rounded PMT shape. 
As discussed in Section~\ref{sec:eff_area}, the \mbox{R5912-100-70} PMT features high sensitivity over the photocathode sphere. This curved design with a constant glass thickness and the optical coupling of the full photocathode area to the pressure housing with silicone elastomer 
ensures the high detection efficiency of the PMT for photons coming from the side.
Sensitivity to short-wavelength (UV) photons is essential for detecting Cherenkov radiation. 
A new glass compound was developed by Okamoto Glass\footnote{\url{http://www.okamoto-glass.co.jp/eng/}} to achieve high UV transmittance and low radioactive noise. There is a strong correlation between light transmittance for wavelengths shorter than \SI{400}{\nano\metre} and the iron content of the glass. 
The D-Egg vessel has a Fe$_2$O$_3$ content of no more than \SI{0.006}{\percent} by weight. This small Fe$_2$O$_3$ content is important for achieving the target UV detection efficiency. The manufacturer's specifications guarantee a \SI{320}{\nano\meter} light transmittance of \SI{75}{\percent} at a thickness of \SI{10}{\milli\metre} after reflection correction~\cite{Shimizu:2017rcd}. A typical transmittance is shown in Figure~\ref{glass_properties}. 
During vessel production, glass from every batch is sampled for future evaluation. The Fe$_2$O$_3$ content and transmittance of the samples are measured regularly throughout the glass production period. 

Approximately half of the dark noise in the IceCube DOM originates from the glass housing~\cite{icecube_instrumentation, IceCube:2010dpc}. Cherenkov light from the radioactive decay of \isotope[40]{K} significantly contributes to the Poissonian noise component and the non-Poissonian scintillation noise. In the D-Egg glass, the standard K$_2$O material is replaced with Na$_2$O to reduce radioactive noise from \isotope[40]{K}. The amount of K$_2$O in the D-Egg glass compound was measured as \SI{0.014+-0.003}{\percent} via X-ray fluorescence analysis. The radioactive material in the D-Egg glass vessel was also measured {with} a gamma-ray spectrometer~\cite{counting_degg:2021}. 
The \isotope[238]{U}, \isotope[232]{Th}, \isotope[40]{K}
concentrations are measured as \SIlist{214.6+-12.9;137.3+-1.4;2548+-16}{\milli\becquerel/\kilogram}, respectively. 

\begin{figure}[t]
\centering
\includegraphics[width=0.65\textwidth]{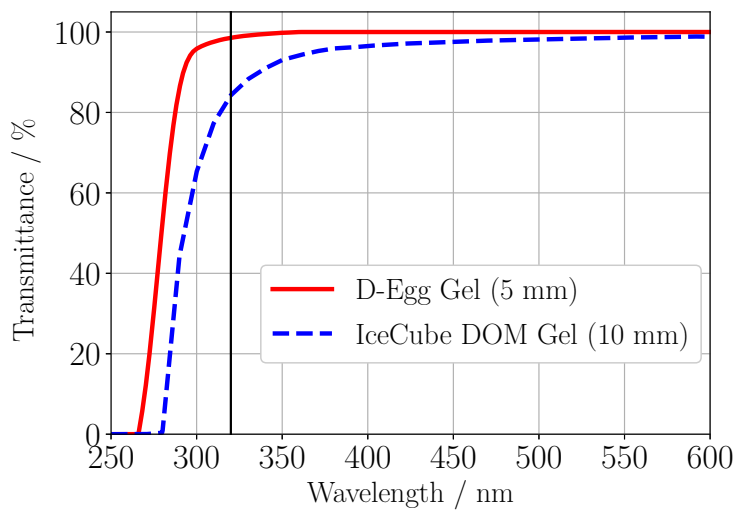}
\caption{Light transmittance of the different silicones as a function of wavelength for those used in the D-Egg (IceCube-Custom silicone) and the IceCube~DOM (RTV6136-D1). Both the selection of the silicone type and the reduction of the thickness contribute to the improvements. 
At the wavelength of \SI{320}{\nano\metre} (shown as a vertical line), 
the transmittance of the D-Egg is \SI{98}{\percent}, while that of the IceCube~DOM is \SI{84}{\percent}.}
\label{fig:gel_properties}
\end{figure}

The PMTs are mechanically attached to the glass housing and optically coupled to it by UV-transmissive silicone. The thickness of the silicone layer is between \SIlist{5.0;6.5}{\milli\metre}, which is approximately half the thickness of the silicone in the IceCube DOM. 
In addition to the light transmittance of the silicone, its hardness is also crucial because it must provide good mechanical support for the PMTs while also being soft and tear-resistant enough to compensate for the shrinkage of the pressure vessel at high pressure during the refreezing phase after deployment. 
Therefore, two custom silicone materials (``\mbox{IceCube-Custom}'' and ``\mbox{IceCube-Custom-HE}'') were developed for the D-Egg by \mbox{Shin-Etsu} Silicone\footnote{\url{http://www.shinetsusilicones.com}} to achieve the best UV transmittance and optimized hardness.
The two materials have the same wavelength dependence in transmittance, exceeding 
\SI{90}{\percent} above \SI{300}{\nano\metre}, which is a significant improvement compared to the optical coupling silicone used in the IceCube~DOM as shown in Figure~\ref{fig:gel_properties}. 
Although they have the same optical properties, \mbox{IceCube-Custom-HE}
has higher elongation and tear strength than \mbox{IceCube-Custom} silicone. 
\mbox{IceCube-Custom} was used to produce the first \num{50}~D-Eggs in the mass production, while \mbox{IceCube-Custom-HE} was used for the remaining \num{260}~D-Eggs.

\begin{figure}[t]
\centering
\includegraphics[width=0.65\textwidth]{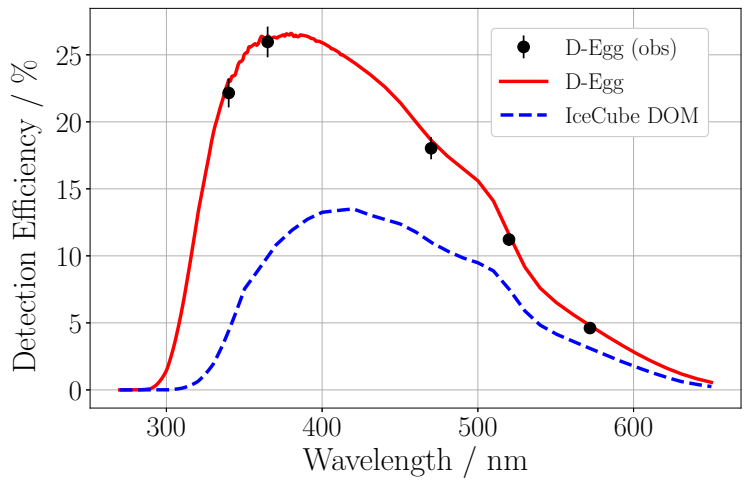}
\caption{Detection efficiency at the threshold of \SI{0.25}{\PE} (photoelectrons) as a function of 
wavelength. The black markers show the values of the glass and silicone transparency measured in the laboratory
convoluted with the averaged PMT QE provided by HPK\@. The red line represents a convolution of the average PMT quantum efficiencies from HPK with the glass and silicone transmittance measured by Okamoto Glass and \mbox{Shin-Etsu} Silicone, respectively, for the D-Egg. The blue line indicates the same quantity for the IceCube DOM as a reference. }
\label{fig:degg_dom_comparison_wl}
\end{figure}

Combining the improved PMT QE with the improved transparency of the silicone and glass, the D-Egg detection efficiency at shorter wavelengths is significantly enhanced compared to that of the IceCube~DOM as shown in Figure~\ref{fig:degg_dom_comparison_wl}.
More details on PMT efficiency calculation are available in Section~\ref{sec:eff_area}.

\subsection{Magnetic Shielding}
The Earth's magnetic field influences the performance of large-area PMTs.
The effect is observed in the magnetic field that is applied parallel to the cathode surface of the PMTs.
The impact of this effect can be reduced by shielding the PMT with high-permeability materials. 
The magnetic shielding for the D-Egg is made of FINEMET$^{\text{\textregistered}}$~\cite{finemet} foil, which is wrapped in a conical shape around the neck of each PMT, similar to that of the Daya Bay optical sensor~\cite{DeVore:2013xma, dayabay}. Because of the costs, ease of integration, and absence of PMT shadowing, this approach was chosen over that for the IceCube~DOMs, which have a mu-metal cage in front of the PMT photocathode~\cite{icecube_instrumentation}.

\subsection{Electronics}

\begin{figure}[t]
\centering
\includegraphics[width=\textwidth]{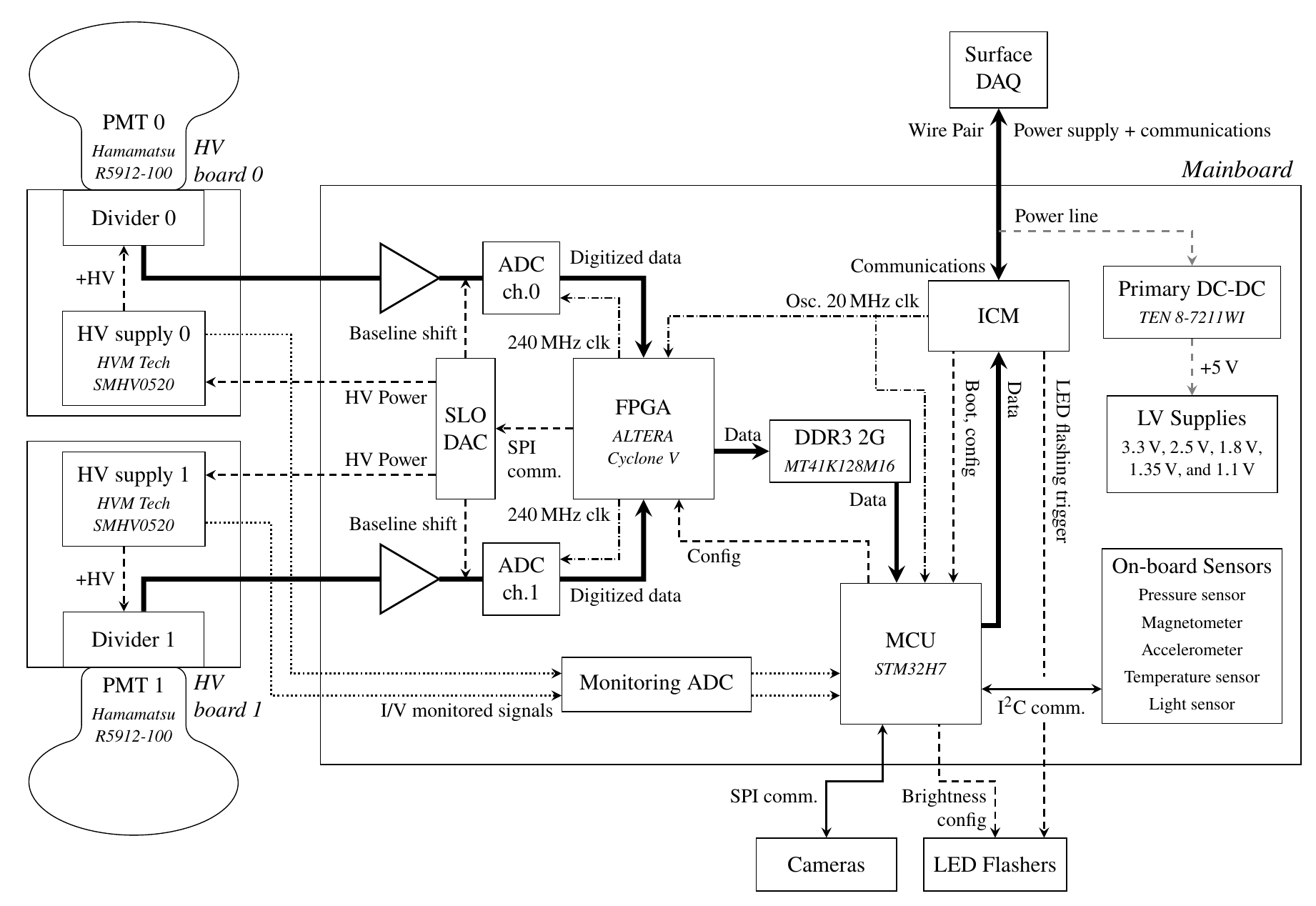} 
\caption{Block diagram of the D-Egg electronics, which consists of the mainboard (Section~\ref{sec:mainboard}), PMT HV boards (Section~\ref{sec:hvboard}), and calibration devices (Section~\ref{sec:calibrationdevices}). 
The mainboard contains an FPGA as the main processor, an MCU running the local DAQ software, and a separate board called the IceCube communications module (ICM) (Section~\ref{sec:icm}). 
The ICM forwards commands from the surface DAQ to the FPGA and the MCU and sends collected data to the surface DAQ system. The ICM also feeds the clock signals {to} all components on the mainboard. 
The MCU {runs} dedicated software to control all systems on the board.  
The FPGA receives digitized data from the ADCs and sends them to the ICM through the MCU. 
The mainboard mounted a \SI{2}{\giga bit} DDR3 SDRAM as a buffer for the digitized waveforms. The FPGA also drives the Slow Control DAC (SLO DAC) to feed the control voltage for the HV generators and to control the baseline of the waveforms by adding an analog offset. 
{The supplied voltage from the surface is stepped down to \SI{+5}{\volt} by the primary DC-DC converter.} 
}
\label{fig:electronics}
\end{figure}

The main tasks of the D-Egg electronics are to enable communication between the surface and the in-ice modules, to control the onboard systems, and to process the signals coming from the PMTs~\cite{IceCube:2008qbc}. 
One of the primary requirements of the D-Egg electronics is the ability to function at temperatures as low as \SI{-40}{\celsius}.
An overview of the D-Egg electronics is shown in Figure~\ref{fig:electronics}. 
The D-Egg mainboard contains custom-designed circuitry to allow for high-frequency signal readout.
A high voltage (HV) board is attached to the base of each PMT. The board mounts a high voltage generator which supplies a controllable PMT operating voltage. The PMT waveforms and the voltage and current monitoring signals from the HV generator are sent to the mainboard through the HV board.
These signals are processed by the mainboard components. 
The IceCube communications module (ICM), which is a small separate board attached to the mainboard, 
is responsible for the communication between the D-Egg and the surface DAQ system. 
A detailed description is provided in the following subsections. 

\subsubsection{Mainboard}\label{sec:mainboard}
The mainboard is the control and data processing center of the D-Egg modules~\cite{Nagai:2019uaz}. It is composed of a \SI{246}{\milli\metre}-diameter circular 14-layer printed circuit board (PCB) containing a large central hole with a diameter of \SI{86}{\milli\metre}. The non-standard board shape is designed to fit inside the space of the glass vessel and around the neck of the 8-inch PMT near the equator of the vessel.
The mainboard contains an onboard high-performance field-programmable gate array (FPGA; ALTERA Cyclone-V 5CEFA5F23I7), a microcontroller unit (MCU; STM32H7), two high-performance low-power analog-to-digital 
converters (ADCs; ADS4149), DC-DC converters, {as well as several environmental sensors}.
The MCU features dedicated software to drive the primary mainboard functionalities, 
configure the FPGA, set the trigger stream, monitor the sensor outputs, and configure and drive the calibration systems. Even after sealing the electronics inside the D-Egg glass vessel, remote updates to the software and firmware of the MCU and FPGA remain possible. 
The mainboard power is provided by the surface system via a pair of wires. The voltage value is within the range of \SIrange{+-60}{+-80}{\volt} and thus above the \SI{+-48}{\volt} present in the current IceCube array in order to increase the available power. The supplied voltage is stepped down to \SI{5}{\volt} by the primary DC-DC converter, and the other necessary voltage lines, such as  \SIlist{3.3;2.5;1.8;1.35;1.1}{\volt} are generated by the secondary DC-DC converters. 

The core functionality of the mainboard is the readout of raw PMT signals. 
Data from each PMT are continuously digitized using a 14-bit ADC with an operation frequency of \SI{240}{\mega\hertz} without any dead time 
after pulse shaping of the analog front-end circuit on the mainboard. 
The pulse-shaping circuit increases the pulse width to improve the charge resolution, particularly for SPE signals. The signal shape is shown in Figure~\ref{fig:spe_wf} and discussed in Section~\ref{sec:gain}. 
The FPGA temporarily stores the digitized data in the buffer in the FPGA and {outputs a signal when the data exceeds a programmable trigger level.}
The outputs are automatically transferred to an external onboard \SI{2}{\giga bit} DDR3 SDRAM (MT41K128M16), which can store hundreds of milliseconds long of the waveforms.  
Several additional data processing, such as data compression or the charge extraction for the waveforms, are performed inside the module in order to remain within the bandwidth limits of the several-kilometer-long main cable. The performed processes are selected depending on the waveform characteristics. 
The FPGA drives the 8-channel Slow Control digital-to-analog converter (SLO DAC). 
The SLO DAC outputs are used as the control voltage for each HV generator and the baseline offset for each readout line. 
The mainboard is equipped with a pressure sensor (LPS22HB), magnetometer (LIS3MDL), accelerometer (ADXL355), temperature sensor (TMP235), and light sensor (ISL76671). These sensors allow monitoring of the internal environment after deployment. 

\subsubsection{PMT High-Voltage Board}\label{sec:hvboard}

\begin{figure}[t]
\centering
\begin{minipage}{.78\textwidth}
\centering
\includegraphics[width=\textwidth]{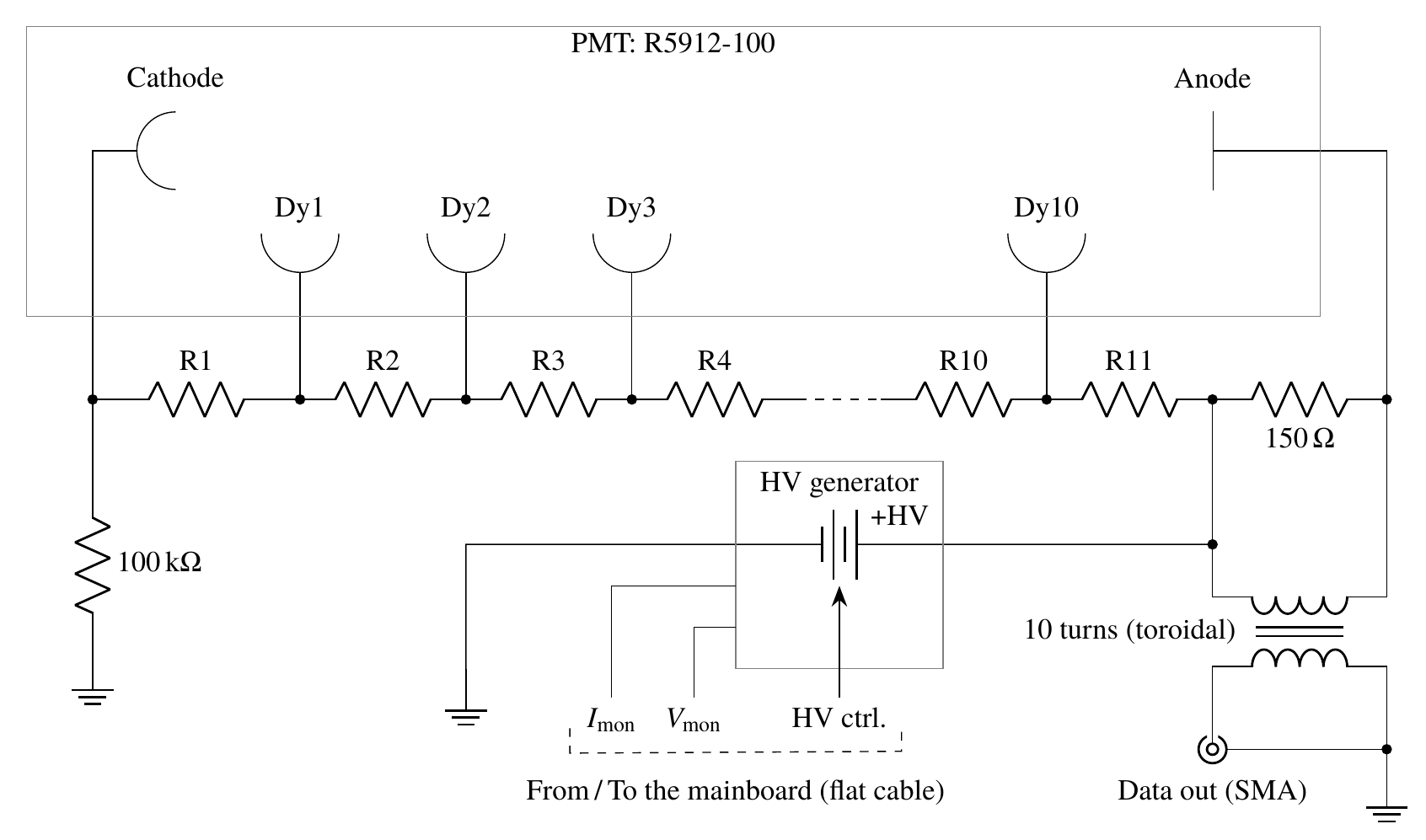}
\end{minipage}\hfill
\begin{minipage}{.2\textwidth}
\centering
\small
\begin{tabular}{|c|c|}
\hline 
Resistor & [\si{\mega\ohm}] \\ \hline 
R1 & 87.6 \\
R2 & 16.8 \\
R3 & 24.6 \\
R4 & 16.3 \\
R5 & 8.34 \\
R6 & 5 \\
R7 & 6 \\
R8 & 7.5 \\
R9 & 11 \\
R10 & 15 \\
R11 & 12 \\ \hline 
\end{tabular}
\end{minipage}
\caption{Schematic view of the PMT HV board for the D-Egg. The decoupling capacitors and damping resistors are not shown for simplicity. The voltage divider has a total resistance of \SI{220}{\mega\ohm}. The HV is generated by a \emph{HVM Technology} HV generator (SMHV0520). The control voltage (range: 0--\SI{5}{\volt}) to apply the high voltage is provided by the mainboard. The values of resistances are summarized in the right table. {Signal coupling from the PMT DC voltage is achieved using a $1:1$ bifilar-wound toroidal transformer.} 
}
\label{fig:HVschematic}
\end{figure}

High voltage is supplied to each PMT from an individual PMT base board containing an HV generator from \emph{HVM Technology}\footnote{\url{https://www.hvmtech.com/}} (SMHV0520), an HV divider circuit, a toroidal transformer, and an analog readout connector. Figure~\ref{fig:HVschematic} shows {the} simplified schematic of the board. The high voltage is derived from the control voltage sourced from the D-Egg mainboard. The two-layer circular board with a diameter of \SI{125}{\milli\metre} is connected to the PMT via a custom-made socket, which supplies a high voltage to the PMT dynodes through an onboard voltage-divider circuit, with ratios recommended by the manufacturer (shown in the table in Figure~\ref{fig:HVschematic}). 

To reduce the risk of electric discharge from the PMT's large photocathode, 
the PMT is operated at a high positive voltage with the anode at a high potential. 
Because of the high voltage applied to the PMT anode, {a $1:1$ bifilar-wound toroidal transformer} is used to decouple the sensitive front-end circuit from the anode. 
%
%
{Transformer coupling provides a lower risk than the commonly-used capacitive coupling, as there is little stored energy available that could possibly damage the sensitive front-end circuitry downstream.}
The transformer acts as a differential circuit that deforms 
the signals from the PMT. 
The toroid with {a TDK H5C2T31x8x19 toroidal core and \num{10}~turns of windings} behaves like a high-pass filter with an exponential impulse response function with a $\sim \SI{5}{\micro\second}$ time constant.
Individual values of this deformation were measured for all boards prior to the installation in the D-Eggs. 
The parameters at the temperature range of \SIrange{-60}{0}{\celsius} are used at the waveform unfolding~\cite{IceCube:2010dpc, IceCube:2013dkx}.

\subsubsection{IceCube Communications Module}\label{sec:icm}

The IceCube communications module (ICM) is a communication interface device for the IceCube-Upgrade modules that is used to transmit information packets between all the modules and the FieldHubs at the surface. 
It is a \SI{35x65}{\milli\metre} rectangular 8-layer PCB mounted on the mainboard. 
It contains a Xilinx Spartan-7 FPGA XC7S25 to control the data transfer between the D-Eggs and the surface DAQ and a \SI{20}{\mega\hertz} oscillator to feed clock signals to the onboard FPGA and the mainboard. 
The communication signals between the D-Egg and the surface DAQ are superimposed onto the DC power line of the mainboard. The signals from the surfaces are separated from the DC current on the mainboard and forwarded to the ICM. 
The signal frequency is \SI{2}{\mega\hertz} with a maximum usable bandwidth of $\sim\SI{1.5}{\mega\bitpersecond}$.
%
Individual D-Egg oscillator timestamps are translated to the central clock domain on the surface using the Reciprocal Active Pulsing (RAPCal) method, as in IceCube~\cite{IceCube:2008qbc}.
%
{The ICM also detects the D-Egg's wire pair address, which is set by jumpers in the external cabling. Hardware interlock signals are also managed by the ICM in response to specific experiment control commands from the surface. This provides a positive assurance against the accidental operation of light emitters, high voltage systems, and reprogramming functions.}
A flashing trigger for the LED flasher system is also provided by the ICM. 
Each ICM has a permanent firmware with hardware write protection that provides basic functionalities. 
Updated firmware may be written to the onboard flash memory and booted from the permanent firmware. 
\subsubsection{Power Consumption}

Since the power to operate the D-Eggs in the ice must be generated on-site at the South Pole, efficient modules are important to minimize operating costs and environmental impact.
The maximum consumption of the optical modules is limited by the resistance losses on each pair of lines. These are given by the length of the cable to the lowest module, the wire resistance, and the output voltage of the power supply at the surface.
For typical operations where the mainboard is running and the two PMTs are held at high voltage, each module consumes approximately \SI{4.4}{\watt}. This satisfies the power consumption requirement of the IceCube Upgrade array. 
Improvements in low-power electronics have reduced the typical power to the point where the total consumption of a D-Egg with its \num{2.8} times effective photodetection efficiency is only $\sim\SI{20}{\percent}$ higher than that of an IceCube DOM.



%
\subsection{Calibration Subsystems}\label{sec:calibrationdevices}

Each D-Egg is equipped with two types of calibration devices with the main goal of  reducing systematic uncertainties related to the properties of the South Pole ice: three camera modules~\cite{IceCube:2021jfx} and an LED flasher system~\cite{Ayumi-ICRC}. The three cameras are attached to a plastic support ring which is placed in the middle of the lower D-Egg hemisphere. 
Each camera module consists of an optical camera and a dedicated blue LED that illuminates the camera's surroundings.
The primary purpose of the camera system is to document the development of ice formed around each module during and after the deployment phase. Additionally, it is possible to determine the position and orientation of an individual module.

The IceCube collaboration utilizes an LED system to measure 
optical scattering and absorption in ice, which are 
the leading systematic uncertainty of physics analyses~\cite{AARTSEN201373,IceCube:2016umi,IceCube:2020acn}.
The flasher system contains \num{12} Roithner \mbox{XRL-400-5O} UV-LEDs attached to a ring-shaped PCB. Each LED emits UV light pulses with a wavelength of \SI{405+-5}{\nano\metre} and a full width at half maximum duration of approximately \SI{6}{\nano\second}. Eight LEDs are directed horizontally, and four emit light vertically downward direction. The LED pulses can be detected by PMTs in the same as well as in the other modules.
\subsection{Penetrator Cable Assembly and Harness}

The connectors on the mainboard for power supply and addressing are connected to the penetrator cable that passes through the \SI{16.5}{\milli\metre} diameter hole in the D-Egg lower glass shell which is sealed to be air-tight. The physical interface outside the glass is a \num{9}-pin water-tight connector from \mbox{Hydrogroup}\footnote{\url{https://www.hydrogroupsystems.com}}.

An aluminum waistband with rubber gaskets encloses the D-Egg's upper and lower glass hemispheres, as shown in Figure~\ref{fig:tmp_degg} (``Harness''). Although not shown in the figure, three vertical \SI{202}{\centi \metre} long steel ropes with eye loops on both ends are attached to the waistband. The three eye loops at each end are bundled using a shackle. 
The harness system can support a vertical weight load of more than \SI{3.5}{\kilo \newton} and withstand a pressure of \SI{70}{\mega\pascal} as reached during the freeze-in process.

\section{Photon Detection Sensitivity}\label{sec:eff_area}

\begin{figure}[t]
\centering
  \begin{minipage}{0.65\textwidth}
    \centering
    \includegraphics[keepaspectratio,width=\textwidth]{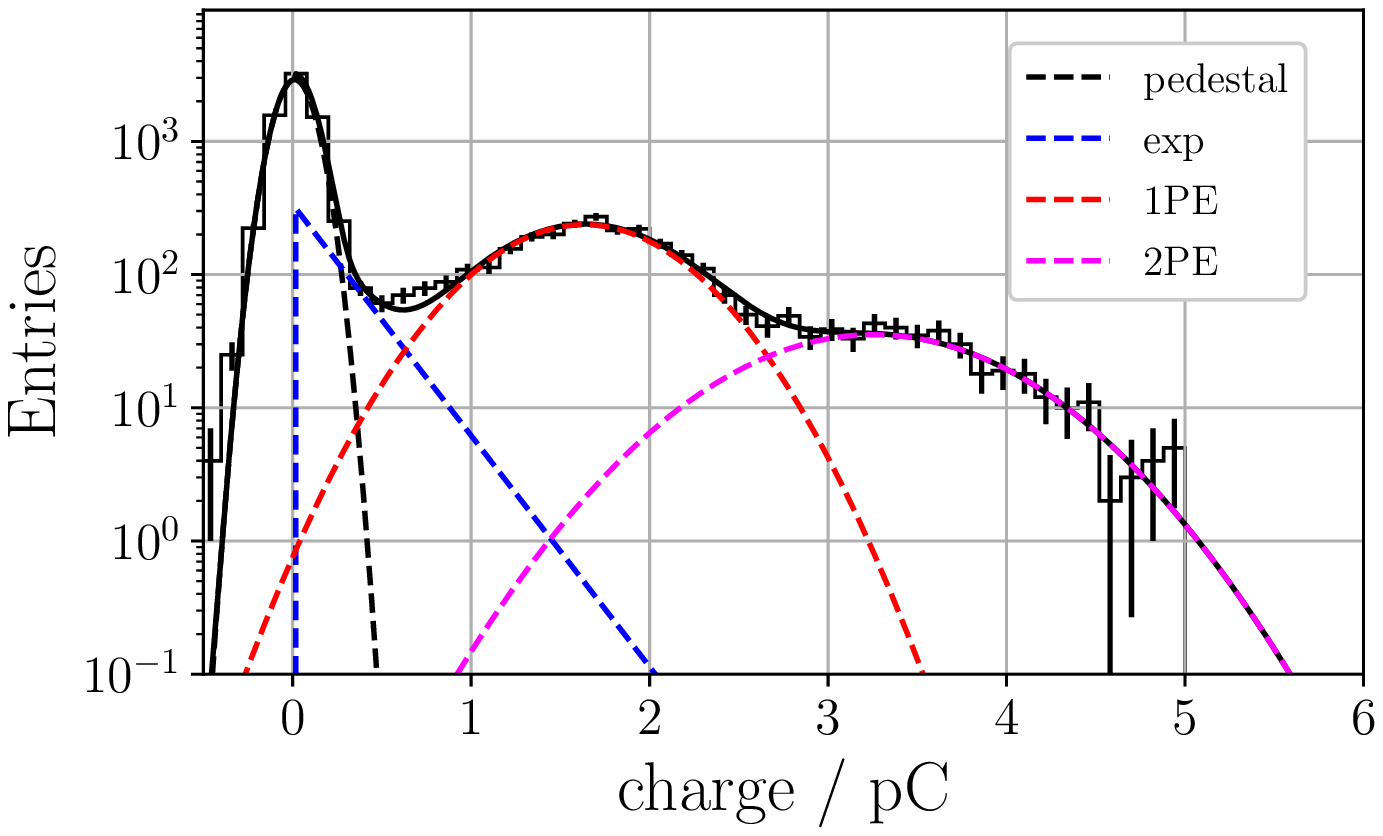}
  \end{minipage}
  \begin{minipage}{0.65\textwidth}
    \centering
     \includegraphics[keepaspectratio,width=\textwidth]{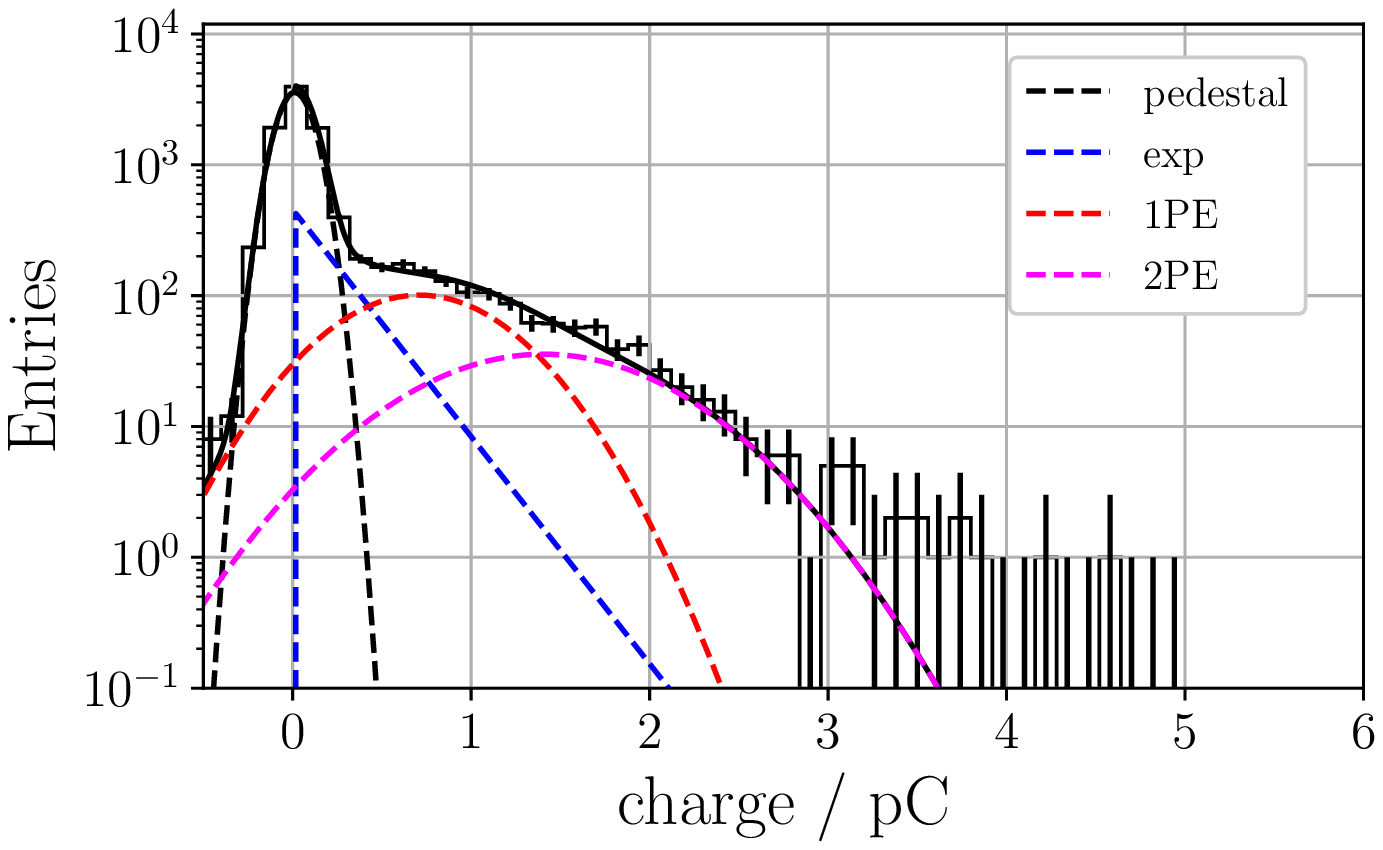}
  \end{minipage}
  \caption{Charge distributions at a center (upper panel) and an edge (lower panel) position of an 8-inch PMT photocathode. The dashed lines are fit terms: pedestal (black), exponential (blue), \SI{1}{\PE} (red), and \SI{2}{\PE} (magenta).}
  \label{SPEfitedge}
\end{figure}

The sensitivity of the D-Egg can be represented by its effective area, 
which is evaluated based on efficiencies of the propagation of the photons to the photocathode and the detection by the PMTs.
The photon propagation efficiency from the module surface to the PMT photocathode was evaluated using GEANT4-based software tools. Photons from a parallel circular beam were directed to the module 
starting in ice with the proper refractive index but without absorption or scattering. The effective area is calculated as follows.
\begin{align}
A(\theta, \phi, \lambda) = \frac{A_0}{N_\mathrm{gen}} \sum_{i: \mathrm{hit}} P(\lambda, \vec{r}_i), 
\end{align}
where $\theta$ and $\phi$ are the zenith and azimuth angles of the beam with 
respect to the axis of the D-Egg,
 $\lambda$ is the wavelength of the generated photons,
 $A_0$ is the area of the circular beam that thoroughly covers the outer diameter of the D-Egg,
 $N_\mathrm{gen}$ denotes the number of photons simulated,
 $P(\lambda, \vec{r_i})$ denotes the detection efficiency
 of the PMT for the wavelength $\lambda$ 
 at the hit position of the $i$-th photon,
 and summation index $i$ runs over photons that 
 hit the photocathode of the PMT.
 $P(\lambda, \vec{r_i})$ was obtained from laboratory measurements: 
the PMT front surface was illuminated by a moving \SI{400}{\nano\metre} wavelength laser spot of a few millimeters in diameter maintaining a normal incidence of the beam.
%

The charges from the anode of PMT were recorded at the PMT output.
The most likely amplification of a photoelectron in the PMT dynode 
system is defined as the single photoelectron (SPE) response of the PMT. 
As can be seen in the upper panel of Figure~\ref{SPEfitedge}, 
this can be measured as a mode in a histogram 
of charge distribution, which is a superposition of pedestal~(\SI{0}{\PE}), SPE~(\SI{1}{\PE}), 
and multiple photons~($\geq$ \SI{2}{\PE}). 
A sum of exponential function and three Gaussian functions 
were fitted to the charge distributions~\cite{BELLAMY1994468}:
\begin{multline}
\frac{\mathrm{d}N(q)}{\mathrm{d}q}  =  \frac{N_0}{\sqrt{2\pi\smash[b]{\sigma_\mathrm{ped}^2}}} \exp\left(-\frac{q^2}{2\smash[b]{\sigma_\mathrm{ped}^2}}\right) 
+ N_1\left[\vphantom{\frac{\epsilon}{\sqrt{\smash[b]{2\pi(\sqrt{2}\sigma)^2}}}}\frac{\alpha}{\tau} \exp\left(-\frac{q}{\tau}\right) + \frac{1-\alpha}{\sqrt{2\pi\sigma^2}} \exp\left(-\frac{(q-q_0)^2}{2\sigma^2}\right) \right. \\ \left.  + \frac{\epsilon}{\sqrt{\smash[b]{2\pi(\sqrt{2}\sigma)^2}}} \exp\left(-\frac{(q-2q_0)^2}{2(\sqrt{2}\sigma)^2}\right)\right], 
\end{multline}
where $q$ is the observed charge; $N_0$, $N_1$, and $\epsilon$ are normalization constants; $\sigma_\mathrm{ped}$ is the width of the pedestal distribution representing no photon hits; $\tau$ is a parameter used to determine the slope of the exponential term representing imperfect amplification by the PMT dynodes; $\alpha$ is the relative contribution of the exponential term with respect to Gaussian terms; $\sigma$ is the width of the SPE distribution; $q_0$ is the most probable charge of an amplified SPE. 
\begin{figure}[t]
\centering
  \begin{minipage}[h]{0.45\textwidth}
      \includegraphics[keepaspectratio,width=\textwidth]{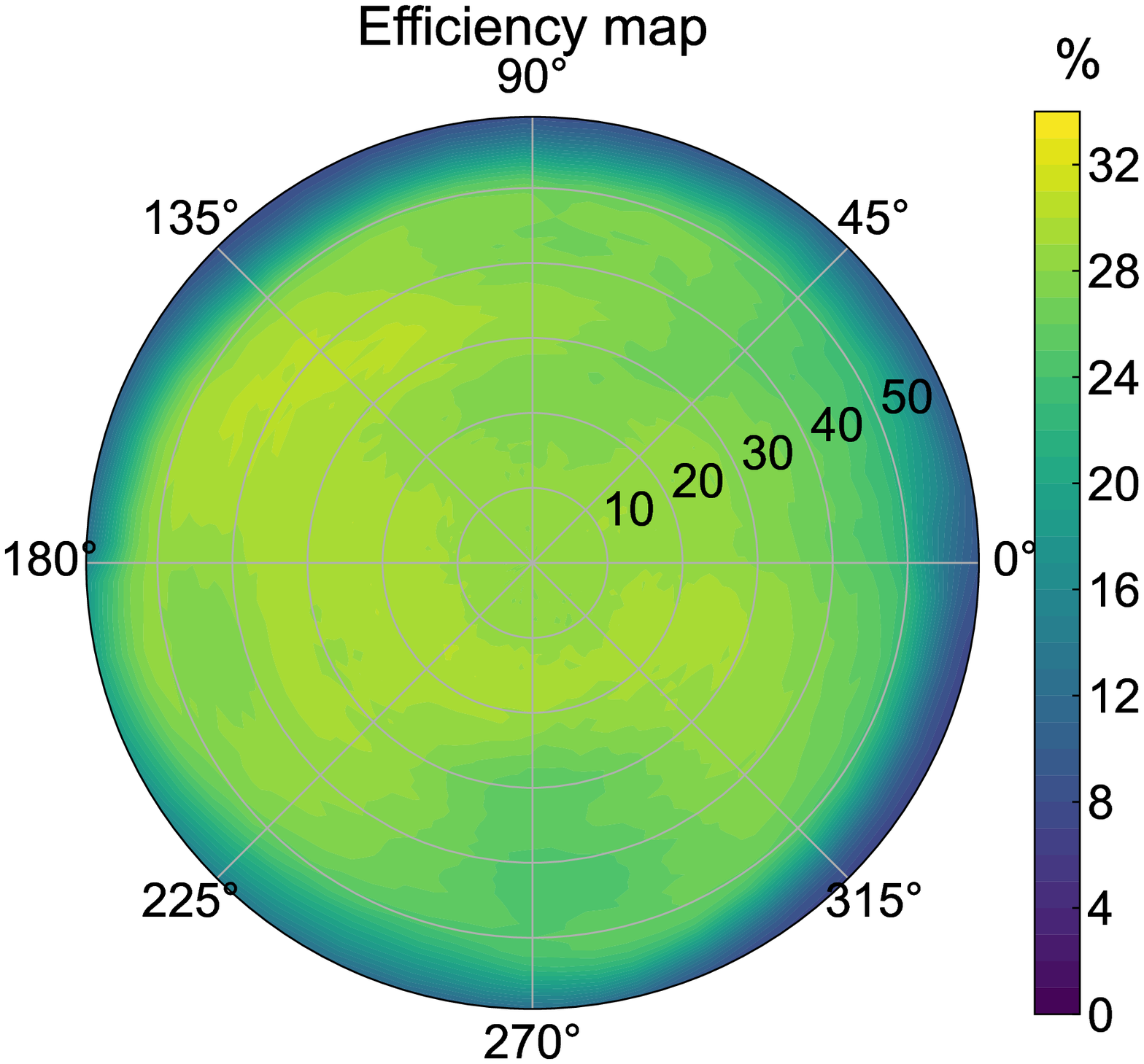}
     \end{minipage}~
  \begin{minipage}[h]{0.54\textwidth}
  \centering
      \includegraphics[keepaspectratio,width=\textwidth]{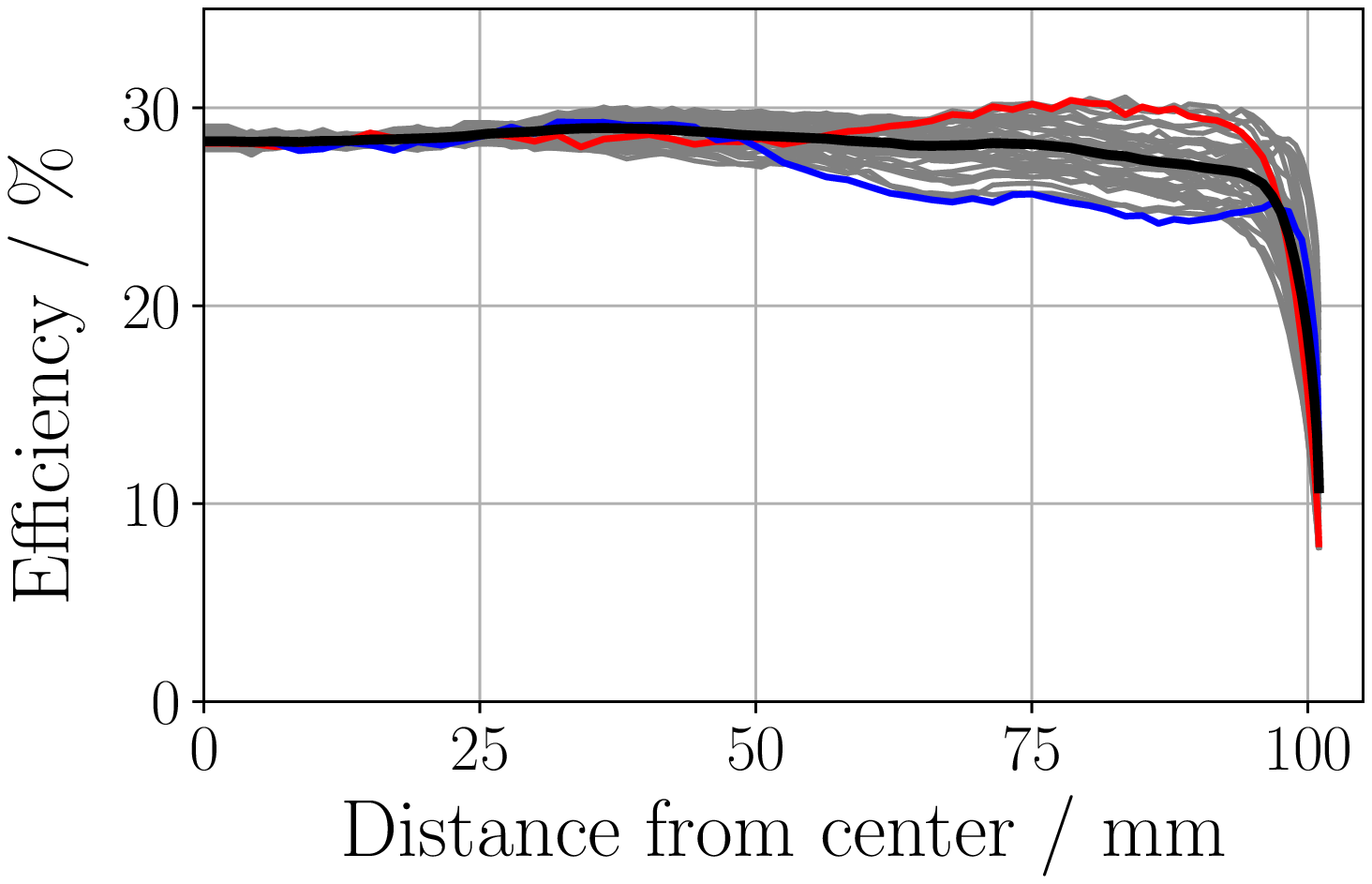}
  \end{minipage}
  \caption{(Left panel) 2D detection efficiency distribution at \SI{400}{\nano\metre} as a function of the hit positions at 
    a threshold of \SI{0.25}{\PE}. 
    The radial axis is the polar angle from the center of the photocathode of the PMT. (Right panel) The same efficiency as a function of distance from the center. Different curves refer to different azimuth angles, and the black line represents the average over azimuth angles. 
    Red and blue lines represent the azimuthal directions showing 
    the maximum and minimum efficiencies. Both figures represent the averages of nine PMTs.}
  \label{MPEscan2d}
\end{figure}
The fit was performed at a large number of positions across the PMT photocathode. 
Figure~\ref{SPEfitedge} shows examples of fits at the center and the edge. The degradation of the amplification efficiency at the edge is clearly visible. 
The single-photoelectron contribution was extracted as follows~\cite{IceCube:2010dpc}:
\begin{align}
\frac{\mathrm{d}N_\mathrm{SPE}(\vec{r},q)}{\mathrm{d}q} = \frac{\alpha}{\tau} \exp \left(-\frac{q}{\tau}\right) + \frac{1-\alpha}{\sqrt{2\pi\sigma^2}} \exp\left(-\frac{(q-q_0)^2}{2\sigma^2}\right)
\end{align}
where the efficiency is defined as:
\begin{align}
P(\lambda, \vec{r}) = \mathrm{QE}(\lambda) \cdot \int_{q_\mathrm{th}}^{\infty}\mathrm{d}q\,\frac{\mathrm{d}N_\mathrm{SPE}(\vec{r})}{\mathrm{d}q}.
\end{align}
Figure~\ref{MPEscan2d} presents the average detection efficiency $P(\lambda, \vec{r})$, evaluated for nine sampled PMTs with a threshold of $q_\mathrm{th}=\SI{0.25}{\PE}$, the standard discriminator threshold used in the IceCube operations, parameterized by the polar and azimuthal angles defined for the axis of the PMT and the center of the photocathode, and as a function of the radius of the cylindrical coordinate system, i.e., the distance from the axis of the PMT in the plane perpendicular to the axis. 
The efficiency represents the performance when IceCube optical modules work as photon counting detectors.
In the following, the PMT detection efficiency is defined as the efficiency measured for a trigger threshold set to \SI{0.25}{\PE} throughout the paper.

For small distances, the \emph{Cherenkov-averaged} sensitivity 
can be used as another benchmark, which was modeled as:
\begin{align}
\bar{A}(\theta, \phi) = \frac{\displaystyle  \int_{\SI{270}{\nano\metre}}^{\SI{700}{\nano\metre}}  \mathrm{d}\lambda~A(\theta, \phi, \lambda)  \mathcal{P}(\lambda) }{\displaystyle  \int_{\SI{270}{\nano\metre}}^{\SI{700}{\nano\metre}} \mathrm{d}\lambda~\mathcal{P}(\lambda) }\quad \text{with} \quad   
\mathcal{P}(\lambda) = \frac{2\pi \alpha}{\lambda^2} \left(1-\frac{1}{\beta^2\, n(\lambda)^2} \right),
\end{align}
where $\mathcal{P}(\lambda)$ is the Cherenkov spectrum, $n$ is the refractive index of the ice, and $\beta$ is the velocity of the traversing charged particles, which is assumed to be \num{1}.
 The absolute magnitude is arbitrary because $\mathcal{P}(\lambda)$ diverges for small wavelengths, 
 and $\bar{A}(\theta, \phi)$ depends on the lower cutoff of $\lambda $.
 
\begin{figure}[t]
\centering
\includegraphics[width=\textwidth]{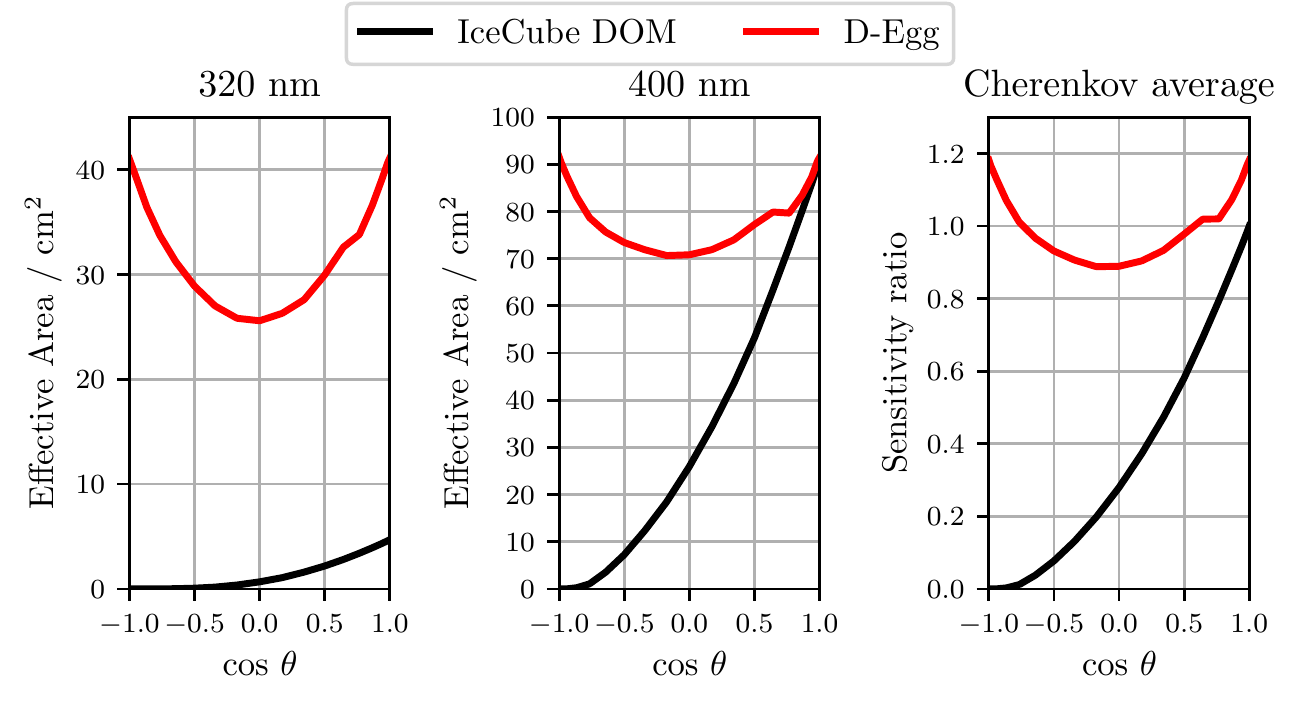}
\caption{Effective area comparison between the D-Egg and the IceCube~DOM as a function of the cosine of the zenith angle of the incident photons at \SI{320}{\nano\metre} (left), at \SI{400}{\nano\metre} (center), and the Cherenkov-averaged sensitivity ratio (right). The downward-facing direction is defined as $\theta=0$ ($\cos{\theta}=1$).}
\label{eff_area}
\end{figure}

 Figure~\ref{eff_area} shows the effective areas of the D-Egg and IceCube~DOM 
 as a function of the cosine of the zenith, where $\theta=0$ is considered the downward {facing} direction. 
 With the two PMTs pointing in opposite directions, and their surface curvature matching their housing glass curvature, the D-Egg features a more homogeneous sensitivity to photons than an IceCube~DOM.
Table~\ref{eff_area_comparison} summarizes the effective areas for the D-Egg and IceCube~DOM after averaging over $\cos{\theta}$ and $\phi$. The Cherenkov-averaged values are shown with respect to those of the IceCube DOM.
The D-Egg sensitivities at \SI{320}{\nano\metre}, \SI{400}{\nano\metre}, and Cherenkov-averaged are higher than those of IceCube DOM by \numlist{24;2.4;2.8} times, respectively. 
The large increase in Cherenkov-averaged sensitivity results from 
improved transmittance of UV light in silicone and glass, and the quantum efficiency of the PMT.

\renewcommand{\arraystretch}{1.25}
\begin{table}
\centering
\caption{Effective areas for the optical modules.${}^*$}\label{eff_area_comparison}
\centering
  \begin{tabular}{|c|c|c|}
  \hline 
   &\ \  IceCube DOM\ \  & D-Egg \\
   \hline
   Effective area (\SI{320}{\nano\metre}) & \SI{1.3}{\centi\metre\squared} & \SI{31}{\centi\metre\squared} \\
   Effective area (\SI{400}{\nano\metre}) & \SI{32}{\centi\metre\squared} & \SI{77}{\centi\metre\squared} \\
   \ \ Cherenkov-averaged sensitivity\ \  & \multirow{2}{*}{\num{1}} & \multirow{2}{*}{\num{2.8}} \\ \ {\small (Ratio to IceCube DOM)} & & \\ 
   \hline 
   \multicolumn{3}{c}{\scriptsize $^*$~An efficiency due to the threshold of \SI{0.25}{\PE} is included in the detection efficiency of PMTs.} 
  \end{tabular}
\end{table}
\renewcommand{\arraystretch}{1}



\section{Acceptance Testing and Verification of Production Modules}\label{sec:degg_performance}


To assess the reliability and expected performance of the D-Eggs at the South Pole, all modules underwent \num{20}~days of continuous testing at various temperatures, including \num{10}~days at an ambient temperature of \SI{-40}{\celsius}. This includes two instances of cooling from room temperature to \SI{-40}{\celsius}, as the region of the South Pole ice in which the D-Eggs will be deployed has temperatures ranging between \SIlist{-10;-40}{\celsius}. Transitions between room temperature and cold temperature simulate variations experienced during transport, storage, and deployment. Extensive testing before shipping is critical because of the high transport costs and the impossibility of repairing modules once they are frozen in the ice. These performance tests focused primarily on the D-Egg PMTs, including the SPE response, dark noise rate, linearity, and timing resolution. Data acquisition is handled exclusively by the D-Egg mainboard with the exception of oscilloscope-based measurements, which are provided for comparison.

\subsection{Setup and Testing Facility}\label{section:setup_facility}

Each module was installed in an individual light-tight box to isolate individual D-Eggs from one another while providing an external light source. A support structure for the boxes was constructed inside a large freezer capable of reaching \SI{-60}{\celsius}. The dimensions of this freezer allowed \num{16} D-Eggs to be tested simultaneously. Temperature and humidity were monitored at several locations {across} the freezer. Power and communications are routed over insulated cables to a set of centralized communication hubs outside. As discussed in Section~\ref{sec:icm}, communication between the D-Eggs and the DAQ is managed by the ICMs. 

\begin{figure}[t]
\centering
\includegraphics[width=\textwidth]{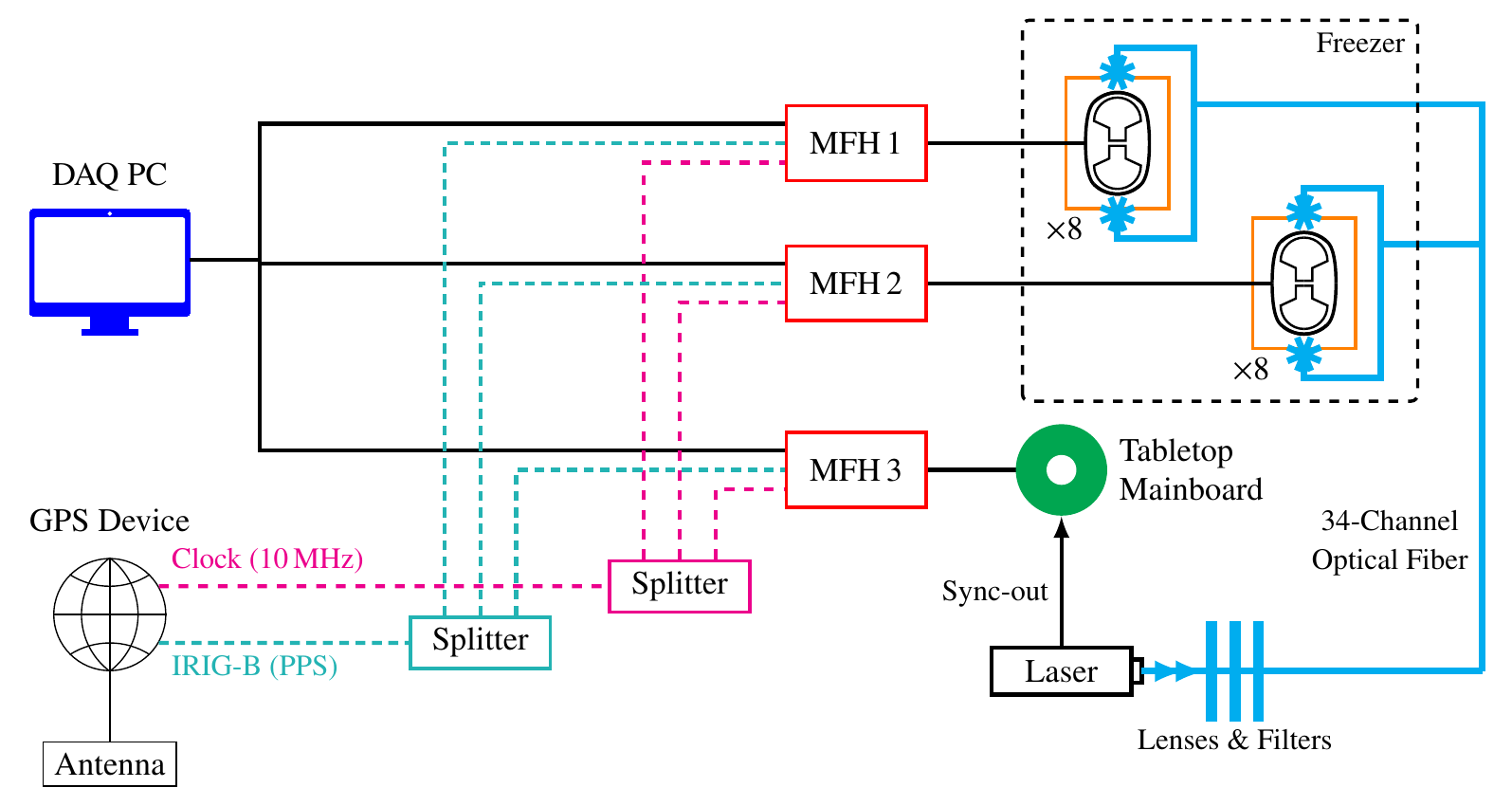}
\caption{Block diagram of the testing setup. The D-Eggs were located inside {individual light-tight boxes} inside the freezer. The light was supplied to each D-Egg's upper and lower PMT via optical fiber. The D-Eggs were connected to Mini-FieldHubs (MFHs) for power and communications. The MFHs were interfaced with a DAQ PC as well as a GPS system for timing information. The GPS signal was equally distributed to the \num{3}~MFHs. Laser timing information was fed into a tabletop version of the D-Egg mainboard.}
\label{fig:setup}
\end{figure}

A block diagram of the testing facility is shown in Figure~\ref{fig:setup}. The primary light source of the setup is a PLP-10 C10196 with a \SI{400}{\nano\metre} picosecond laser diode head M10306 from HPK\@.\footnote{\url{https://www.hamamatsu.com/resources/pdf/sys/SOCS0003E_PLP-10.pdf}} Light pulses are delivered to the upper and lower PMT of each D-Egg via optical fiber from the outside of the freezer. The D-Eggs are connected to Mini-FieldHubs (MFHs) for power and communications. The MFH is a simplified version of the full-scale FieldHub, which interfaces all modules to a single DAQ PC. A tabletop version of the D-Egg mainboard is connected to the synchronization laser timing output for timing-specific measurements. A \SI{10}{\mega\hertz} clock and the IRIG-B time signals from the GPS are split and fed into each MFH to allow synchronization of the MFH clocks to UTC. The D-Egg internal timestamps are then translated to UTC using the standard RAPCal procedure that connects the D-Egg ICM and MFH clock domains.

 The repetition rate of the light source is controlled using a programmable function generator. The laser light intensity was attenuated by two \num{6}-channel programmable filters and focused by lenses before entering the \num{34}~channels of the optical fiber bundle, which transport light into the freezer for each D-Egg PMT. Light levels ranging from low-occupancy SPE to more than \SI{200}{\PE} were available to test the PMT response. The spread in intensity between the fiber channels was measured to be \SI{15}{\percent}. The trigger time of each laser pulse (via sync-out) was independently digitized and recorded using an additional synchronized ICM mounted on a tabletop D-Egg mainboard.

\subsection{Gain Calibration}\label{sec:gain}

\begin{figure}[t]
\centering
\includegraphics[width=0.8\textwidth]{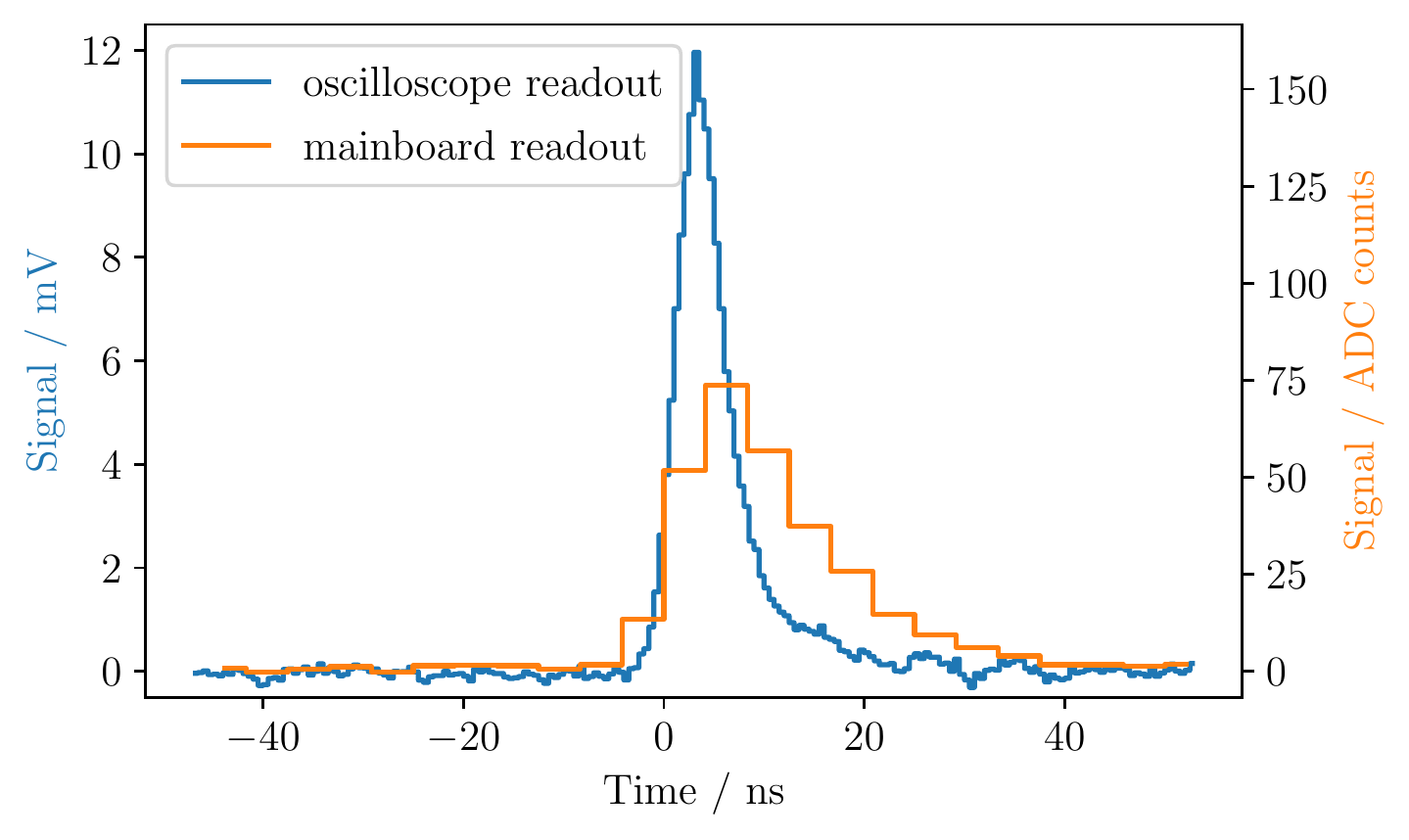}
\caption{Averaged SPE waveform at a gain of \num{e7}. The analog PMT pulse is shaped by the mainboard electronics, and the amplitude is inverted from negative to positive. Oscilloscope data has also been inverted for comparison.}
\label{fig:spe_wf}
\end{figure}

The IceCube Upgrade will operate the D-Egg PMTs at a gain of \num{e7}, with most PMTs requiring approximately \SI{1500}{\volt} applied to the cathode. Calibration to determine the operational high voltages can be performed using SPE-level pulses. For a bare PMT, the typical pulse amplitude is approximately \SI{12}{\milli\volt}, with a full-width half-maximum (FWHM) duration of approximately \SI{14}{\nano\second}. Figure~\ref{fig:spe_wf} shows the average SPE waveform for a representative module operating at \num{e7}~gain, as measured by an oscilloscope (\SI{2}{\giga\hertz} bandwidth, \SI{10}{\giga\SPS}) 
and the mainboard using a \SI{2.0}{\milli\volt} trigger threshold. 
The width of the raw PMT pulse is broadened by the mainboard front-end electronics, resulting in a measured SPE pulse height of approximately \SI{6}{\milli\volt} and an FWHM duration of approximately \SI{20}{\nano\second}. Samples before the trigger are pre-buffered by the mainboard to calculate the baseline. The baseline can be adjusted by changing the offset from the SLO DAC. 

A charge distribution was constructed from the SPE waveforms and used to extract the PMT gain (see Figure \ref{SPEfitedge}). The charges were calculated by integrating the waveforms from \num{10}~bins before the waveform peak until \num{15}~bins after the peak (\SIrange{-41.7}{62.5}{\nano\second}). Waveforms were obtained using a threshold of \SI{1.875}{\milli\volt}. A Gaussian was fitted to the charge histogram, where the peak position indicates the PMT gain at a given high voltage. This procedure was repeated for different high voltages until the extracted gain was within \SI{+-2}{\percent} of \num{e7}.

\subsection{Dark Noise}

Dark noise is a type of background not originating from an external photon arriving at the detector. Sources of dark noise include radioactive processes, thermionic cathode emissions, electronic noise, and scintillation within the pressure vessel and PMT glass.
The dominant contributor to radioactive processes is the decay of $^{40}$K in the borosilicate glass pressure vessel enclosing the D-Egg. PMTs that are optically connected to the glass observe a dark noise rate that triggers the DAQ at approximately \SI{3}{\kilo\hertz} in the laboratory or approximately \SI{1}{\kilo\hertz} in ice when operated at an ambient temperature of \SI{-40}{\celsius}. Cooling the PMT reduces thermionic emissions, thereby reducing the overall dark rate of the PMT.

\begin{figure}[t]
\centering
\includegraphics[width=0.8\textwidth]{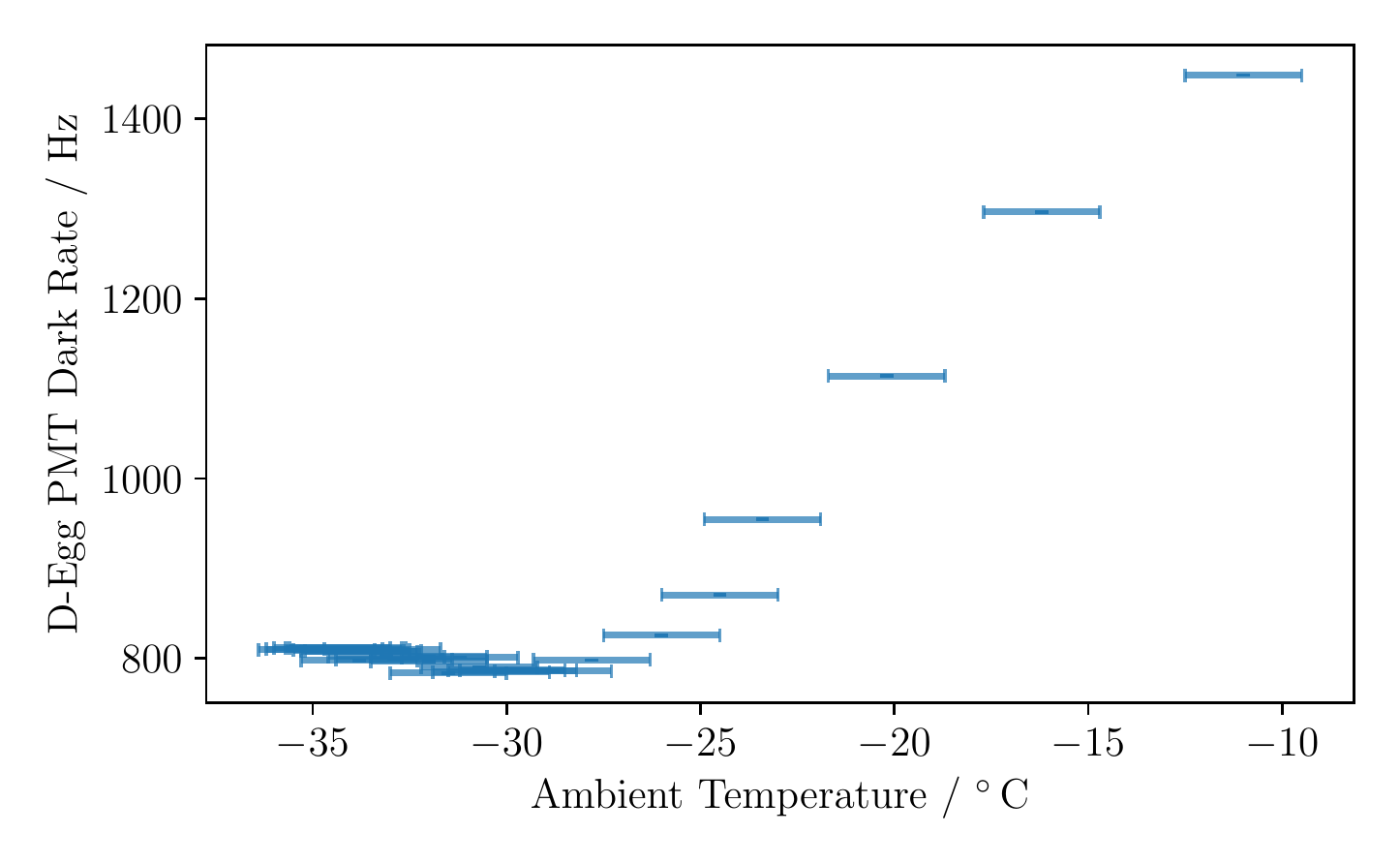}
\caption{Dark rate versus ambient temperature for a representative D-Egg PMT. Each data point was collected at a fixed temperature, gain, and threshold. Ambient temperature is measured with a thermocouple located inside the freezer. Depending on the location of the module and time during the freezer cycle, the measured ambient temperature can vary from the actual module temperature by about \SI{+-1.5}{\celsius}. }
\label{fig:darkrate_temperature}
\end{figure}

This effect is shown in Figure~\ref{fig:darkrate_temperature}, where the dark rate was sampled over a range of temperatures for \SI{1000}{\second} at each temperature. Before measuring at each point, the gain of the PMT was re-calibrated and the appropriate threshold re-calculated. The variation in the temperature of each module is a function of its location inside the freezer as well as the mechanical operations of the freezer. This results in an average temperature difference of \SI{+-1.5}{\celsius}. The ambient freezer temperature was measured using a thermocouple at a fixed location. {Typical operational temperatures at the experimental site in Antarctica have ambient ice temperatures ranging between \SIlist{-40;-20}{\celsius}.}

Once the module is deployed, the refractive indices of the glass (\num{1.49}) and the deep South Pole ice (\num{1.78}~\cite{rice_refractive_index}) match more closely than glass and air (\num{1.00}), leading to an observed drop in dark rate relative to laboratory measurements. This originates from a decrease in internal reflections at the boundary between the glass and the surrounding medium (ice/air). A good approximation for the glass-to-ice boundary for a deployed D-Egg is achieved by covering the outer glass surface with black vinyl tape. This causes outward-going photons from the decays inside the glass to be absorbed rather than reflected back towards the PMT~\cite{rawlins:2001}. Empirically, taping decreased the dark rate by a factor of $\sim\num{2.4}$ as measured at \SI{-40}{\celsius}. This was determined by comparing the dark rates of multiple D-Eggs, measured under both taped and standard conditions, and is applied in Figure~\ref{fig:darkrate_temperature}.

\begin{figure}[t]
\centering
\includegraphics[width=0.7\textwidth]{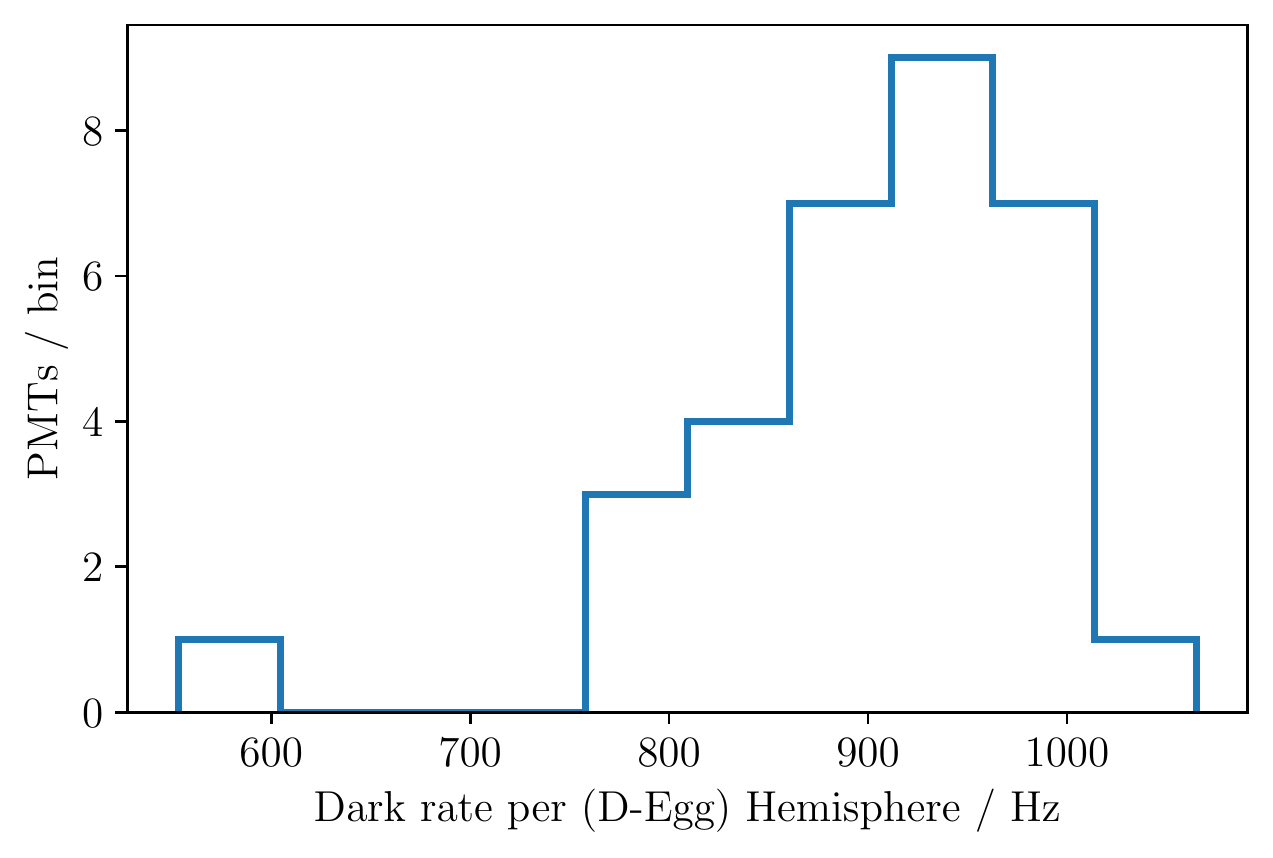}
\caption{Dark rate for \num{32} integrated PMTs measured at an ambient temperature of \SI{-40}{\celsius}. Rates were measured with PMTs operating at \num{e7}~gain, with a threshold of \num{0.25}~times the average SPE peak amplitude, and an artificial dead time of \SI{100}{\nano\second} applied in software. The presented dark rates are scaled down by a factor of \num{2.4}, taking into account the difference in refractive index between air and ice.}
\label{fig:darkrate}
\end{figure}

Figure~\ref{fig:darkrate} shows the reflection-corrected dark rates for \num{32}~PMTs that are fully integrated into modules expected to be deployed in ice at an ambient temperature of \SI{-40}{\celsius}. These events were triggered when the average of two consecutive waveform bins exceeded {a threshold of \SI{0.25}{\PE} where \SI{1}{\PE} is the peak height of the average SPE pulse.} In addition, an artificial dead time of \SI{100}{\nano\second} was applied to prevent the recording of multiple triggers from the same pulse.
The observed mean dark rate of \SI{908}{\hertz} per PMT, or \SI{1.8}{\kilo\hertz} in total, can be compared with that of the IceCube DOM of \SI{870}{\hertz} and that of the DeepCore DOM of \SI{1.2}{\kilo\hertz} taking the empirical reflection-correction factor into account~\cite{deepcore}. Normalizing the D-Egg dark rate to the Cherenkov-averaged effective area of the IceCube DOM results in a rate of \SI{568}{\hertz}.
This indicates that the D-Egg design shows a reasonable improvement in the dark rate.

\begin{figure}[tb]
\centering
\includegraphics[width=0.49\textwidth]{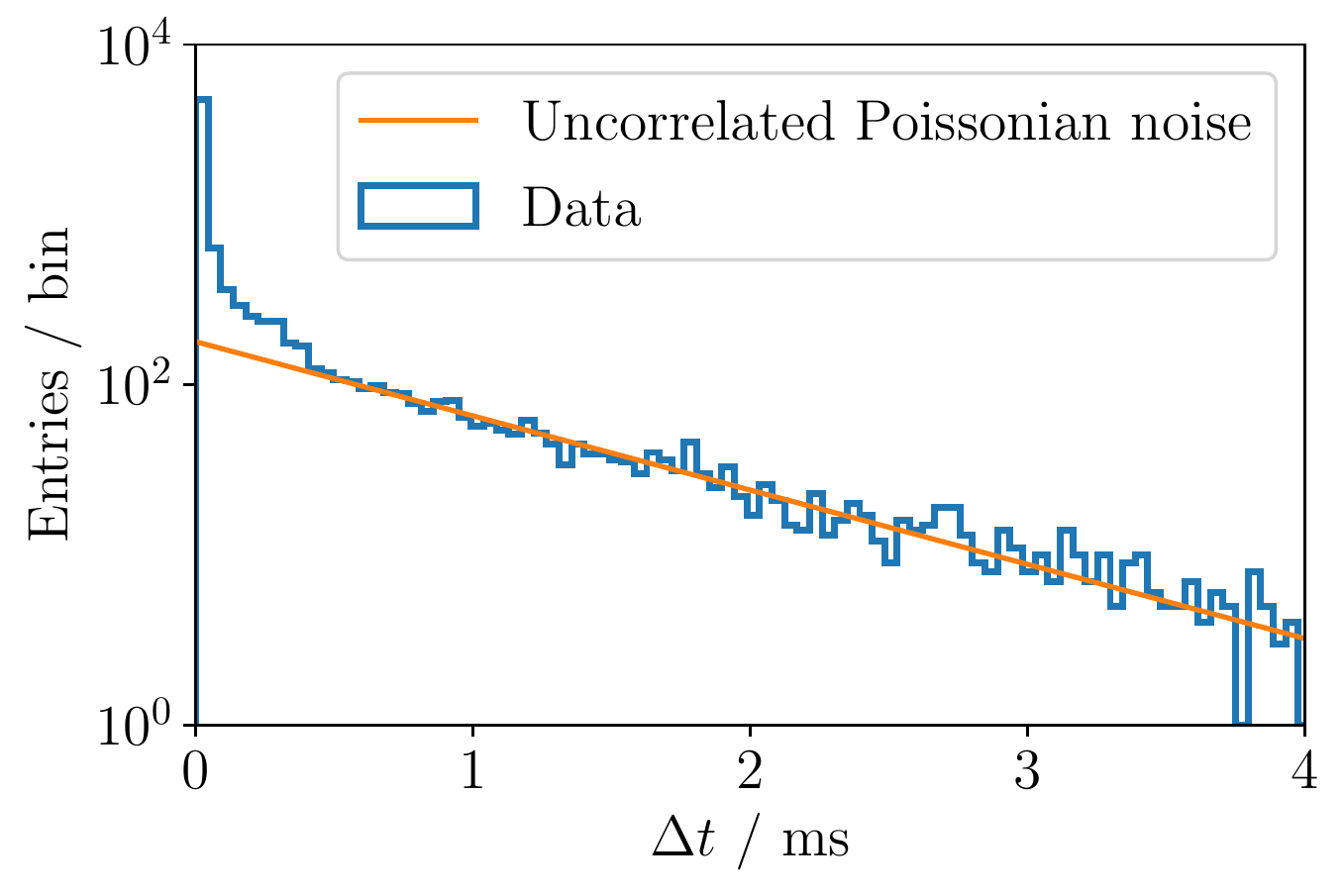}
\includegraphics[width=0.49\textwidth]{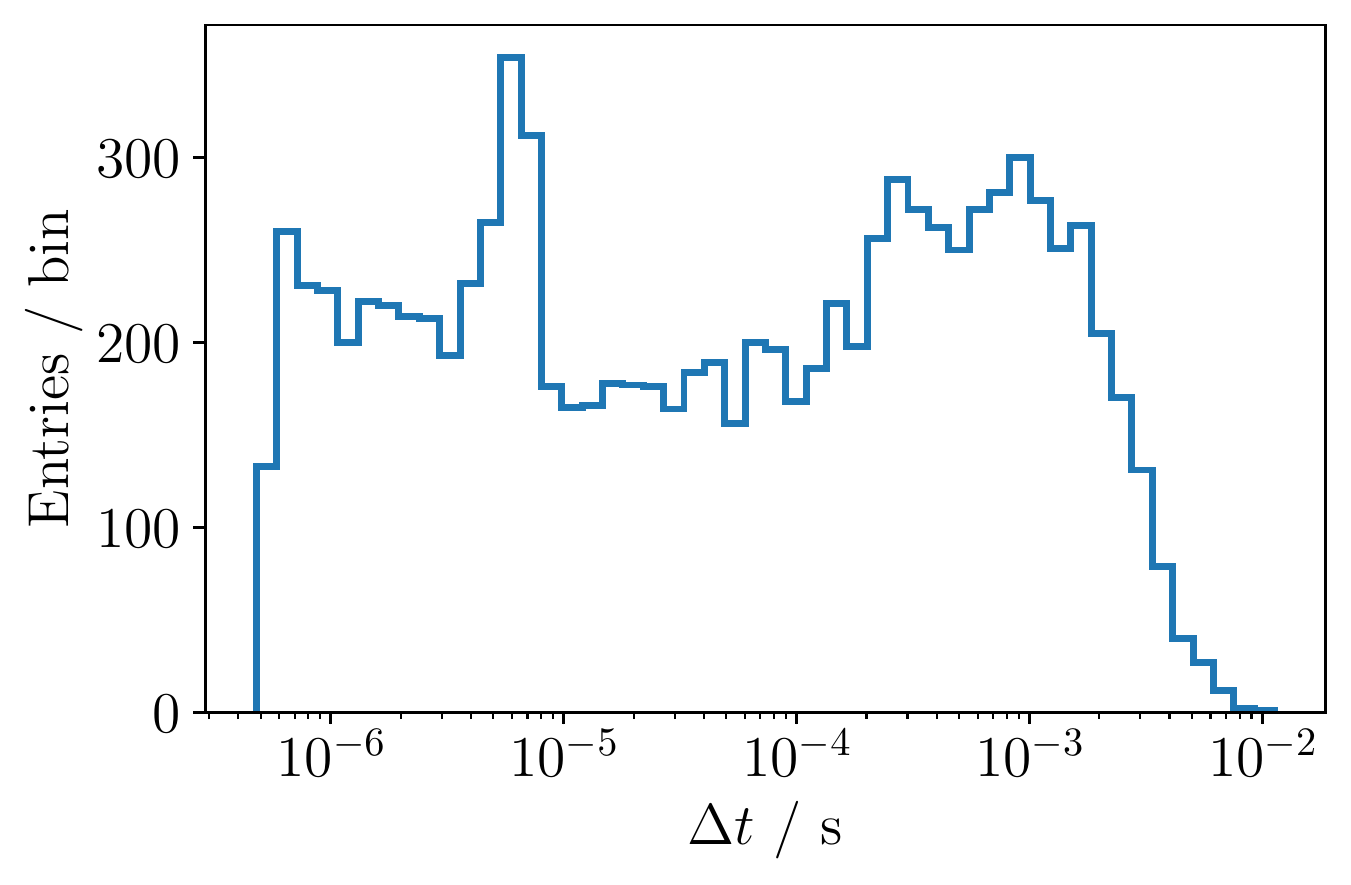}
\caption{Typical $\Delta t$ distribution for a D-Egg PMT. The uncorrelated component of thermal noise hits can be described by a Poisson distribution with a mean of about \SI{1}{\kilo\hertz}. The right side shows the same distribution on a logarithmic scale to highlight dark noise components at shorter time scales, e.g., PMT afterpulses at time scales of around \SI{5}{\micro\second}.}
\label{fig:delta_t}
\end{figure} 

Figure~\ref{fig:delta_t} shows the distribution of the time intervals between consecutive triggers ($\Delta t$). This is particularly important for building a low-threshold in-ice-triggering scheme, as high dark rates can contribute to false coincident triggers during scientific data-taking. The distribution is primarily composed of two components: large time differences originate from uncorrelated noise hits and can be described by a Poisson distribution~\cite{Larson:2013xbf}. 
The second component is the correlated hits, coming from radioactive decays in the pressure vessel and PMT afterpulses. These are visible as an excess over the Poisson expectation at time differences below $\sim\SI{5e-4}{\second}$.

\subsection{Linearity}

Measurements of the PMT linearity as a function of peak current are essential for energy calibration. Linearity was measured as the combined response of the PMT and front-end electronics versus the ideal response for a given light source. Picosecond long laser pulses were directed into each \mbox{D-Egg} dark box via an optical fiber after either being attenuated by nine different settings of neutral density filters with strengths of \SIlist{1;2.5;5;10;12.5;16;25;32;50}{\percent}, or passing through the tenth ``open'' setting (\SI{100}{\percent}).

The laser intensity was tuned such that \SI{2.5}{\percent} transmittance corresponded to a signal of a few tens of photoelectrons observed by the PMT and \SI{100}{\percent} resulting in pulses of at least \SI{200}{\PE}. This ensures that at least one intensity setting is well within the expected linear region. Fluctuations in light intensity of \SI{15}{\percent} were observed at the fiber outputs due to the fiber coupling location at the $1:34$ channel split, transport losses, and geometric effects associated with the fiber support in the dark box.

\begin{figure}[tb]
\centering
\includegraphics[width=0.8\textwidth]{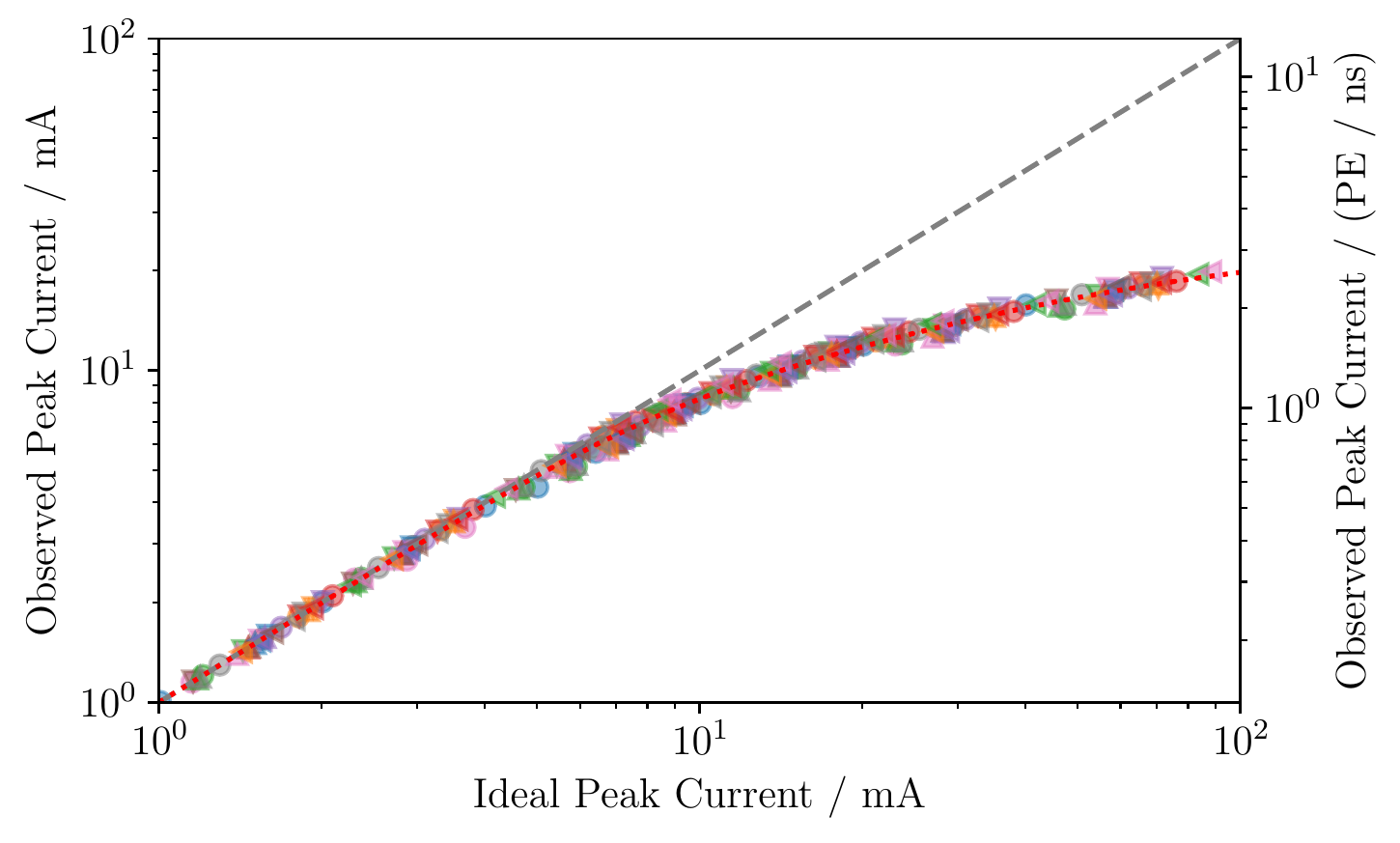}
\caption{Linearity response for \num{32} integrated D-Egg PMTs at \SI{-40}{\celsius}. Each PMT was measured at \num{10} filter transmittance settings, spanning the range of tens to hundreds of PE. The unique mark and color combinations correspond to a single PMT. The spread in ideal peak current is driven mainly by variations in the optical fiber location and angle relative to the PMT. A $1:1$ line (grey) is used to guide the eye between the linear and non-linear region, and the observed peak current described in eq.~\eqref{eq:obs_current} (red) is fit to the data. Divergence from the linear regime at lower ideal peak current values is driven primarily by the broadening behavior of the mainboard electronics.}
\label{fig:linearity_current}
\end{figure}

The repetition rate of the laser emission was maintained at \SI{100}{\hertz} with a fixed pulse width of a few picoseconds. The peak current, shown in Figure~\ref{fig:linearity_current}, is a quantity that is independent of the input pulse width. Triggers from atmospheric muons, at a rate of approximately \SI{10}{\hertz}, were removed using a cut on the fixed repetition rate of the laser.
{We want to note that the linearity of the total charge depends on the input pulse width. The D-Egg is linear up to \SI{60}{\PE} with picosecond-level pulses, but maintains linearity up to \SI{2000}{\PE} for \SI{150}{\nano\second} wide pulses.}

The departure of the system from the linear regime is driven by a combination of the PMT and mainboard front-end signal shaper behavior, specifically the slew-rate limitation. The linearity of the observed peak current $I_{\mathrm{obs}}$ versus the ideal peak current $I_{\mathrm{ideal}}$ can be modeled as:
\begin{equation}
    \label{eq:obs_current}
    I_{\mathrm{obs}} = I_{\mathrm{ideal}} \frac{\ln \left( 1 + \frac{I_{\mathrm{ideal}}}{p_2} \right)}{\ln \left( 1 + \frac{I_{\mathrm{ideal}}}{p_2} \right) + \frac{I_{\mathrm{ideal}}}{p_0} \ln \left( 1 + \left( \frac{I_{\mathrm{ideal}}}{p_1} \right)^3 \right)}
\end{equation}
where the modelling parameters are $p_0 = \SI{36.71}{\milli\ampere}$, $p_1 = \SI{1.89}{\milli\ampere}$, and $p_2 = \SI{0.14}{\milli\ampere}$~\cite{IceCube:2010dpc}.
%


\FloatBarrier

\subsection{Timing Resolution}

Reconstruction in IceCube relies heavily on timing information. For PMTs, the time it takes for a photoelectron emitted from the photocathode to arrive at the anode is known as the transit time. Variations in the transit time between photoelectrons result in a transit time spread (TTS), which heavily affects the timing resolution, although the overall timing capabilities of the \mbox{D-Eggs} are determined by a combination of PMT and mainboard.

\begin{figure}[t]
\centering
\includegraphics[width=0.8\textwidth]{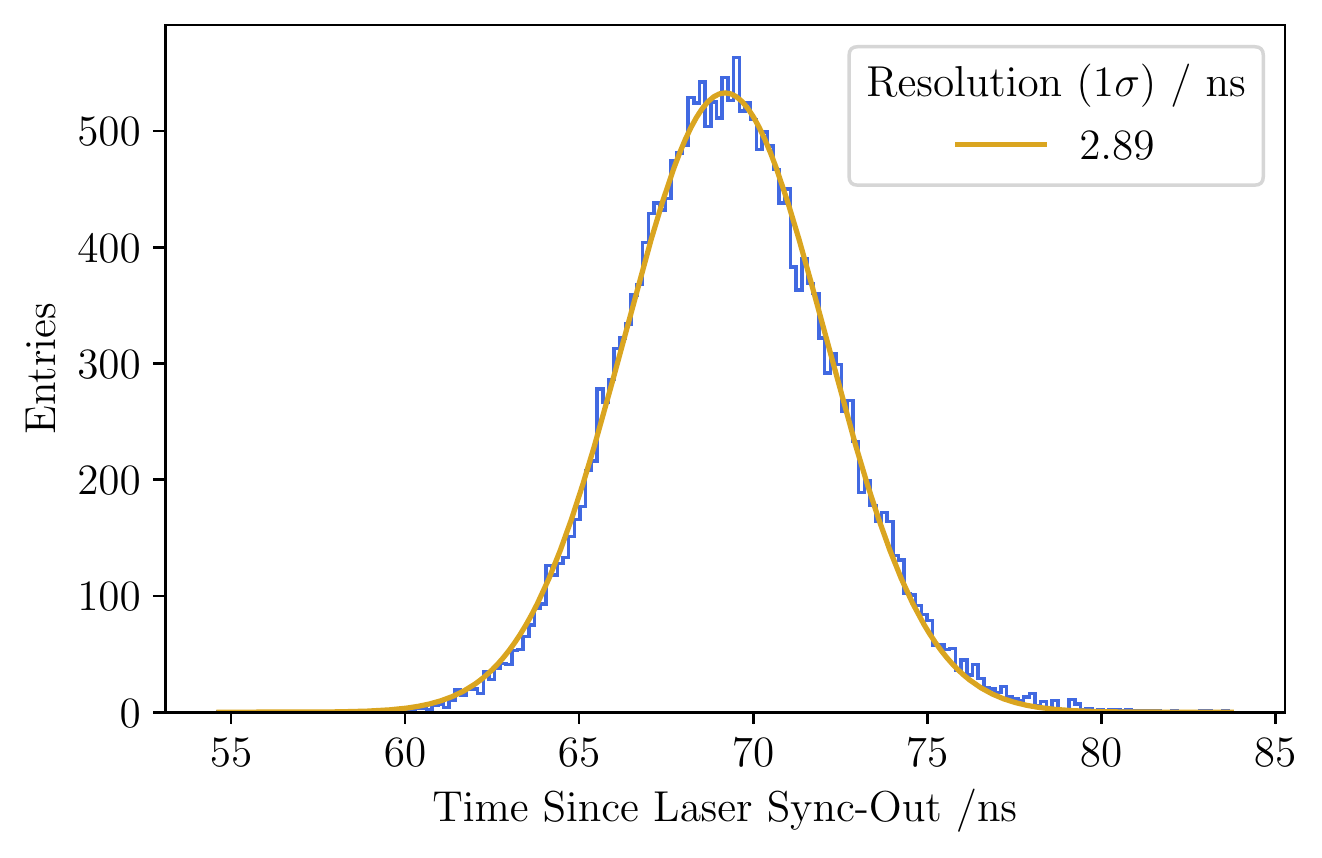}
\caption{D-Egg PMT transit time distribution for a low-occupancy SPE-level light source. The SPTR is extracted as \SI{2.89}{\nano \second} and represents the typical D-Egg PMT performance. The transit time peak position has no impact on the SPTR or TTS of the PMT and is a result of the experimental setup.}
\label{fig:tts}

\end{figure}

\begin{figure}[t]
\centering
\includegraphics[width=0.8\textwidth]{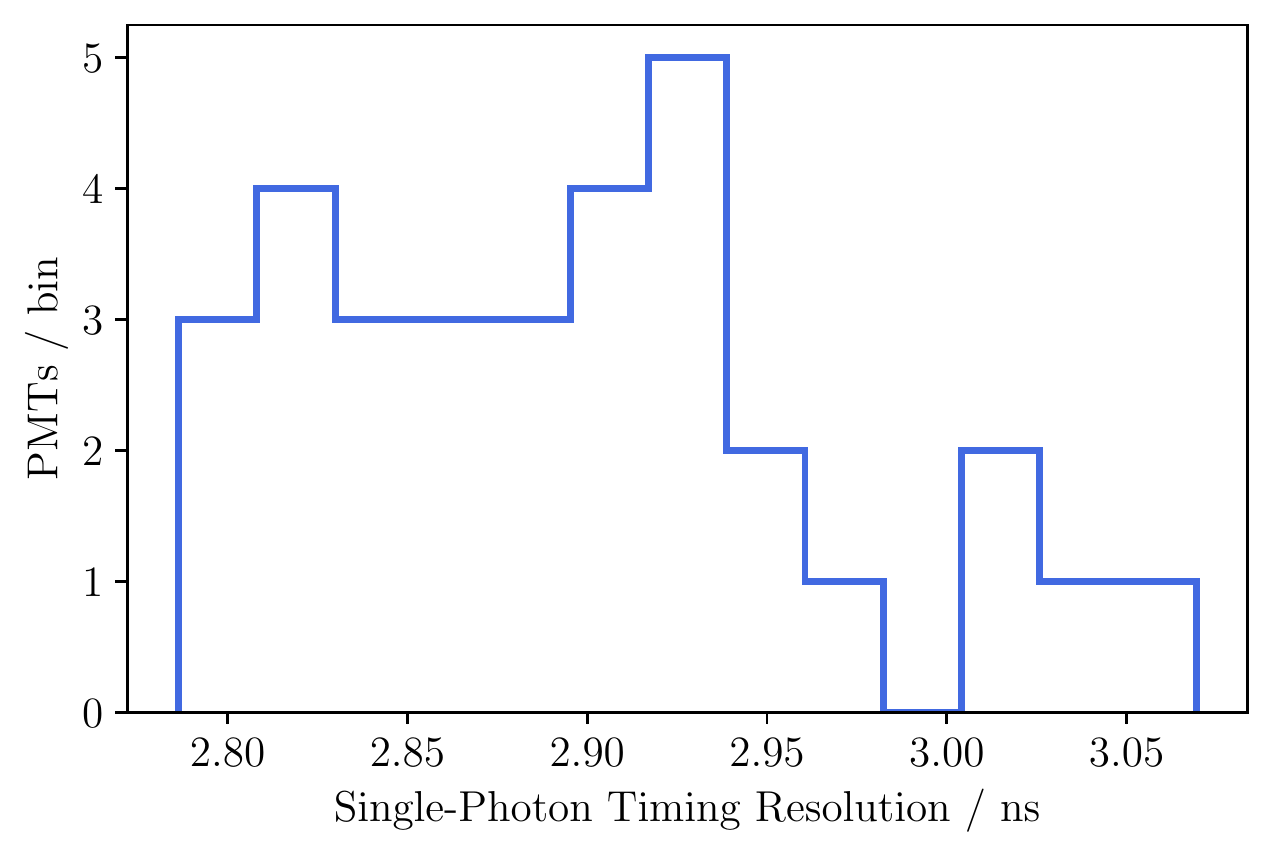}
\caption{D-Egg PMT single-photon timing resolution (SPTR) distribution for 32 modules. All modules satisfy design requirements of having an SPTR of less than \SI{5}{\nano \second}.
}
\label{fig:tts_hist}
\end{figure}

To determine the single-photon timing resolution (SPTR), SPE-level laser pulses were injected into the \mbox{D-Eggs}, and the rising edge times were identified. The timing was determined as discussed in Section~\ref{section:setup_facility}. To preserve nanosecond precision, the ICM clocks on the D-Egg mainboard and Mini-FieldHub were re-synchronized approximately every \SI{1}{\second} using the RAPCal procedure at regular intervals throughout the measurement. The SPTR was determined by fitting the distribution of the differences between the recorded D-Egg mainboard time and the reference mainboard time with a Gaussian function and extracting the width. On average, the \mbox{D-Egg} PMT has a well-defined single-photon timing resolution of the order of \num{2}--\SI{3}{\nano\second}, with bare PMT values measured between \SIlist{1.8;2.0}{\nano \second}. Representative distribution of the SPTR is shown in Figure~\ref{fig:tts}, where one entry corresponds to the difference in time between the pulse from the PMT recorded by the \mbox{D-Egg} electronics and the digitized laser sync-out signal.  The absolute peak position of the distribution depends on the experimental setup but does not influence the measured TTS and SPTR. A Gaussian was fitted to the data, returning a resolution of $\sigma = \SI{2.89}{\nano \second}$. This result as compared to the other measured \mbox{D-Egg} PMTs is displayed in Figure~\ref{fig:tts_hist}.


Design requirements place an upper limit of \SI{5}{\nano \second} on the timing resolution for a given device, which all modules satisfy. For context, in IceCube, the typical spread in leading-edge times for photoelectrons to be recorded from a \SI{10}{\PE} interaction \SI{30}{\metre} away is approximately \SI{3}{\nano \second}, which is similar to the typical D-Egg PMT SPTR. 
\subsection{Double Pulse Feature Extraction}

Waveform-based feature extraction for two multi-PE pulses separated by only a few nanoseconds is a critical tool for identifying high-energy tau neutrino-charged current interactions. A function generator supplied the laser with a trigger to emit two picosecond-wide bursts separated by \SI{20}{\nano\second}, followed by a \SI{1}{\micro\second} delay before the next set of pulses. The intensity of each pulse was configured to be in the linear regime. The time between sets of pulses at this light level ensured high occupancy without saturating the readout system and allowed the waveform to return to the baseline before the next trigger.

\begin{figure}[t]
\centering
\includegraphics[width=0.8\textwidth]{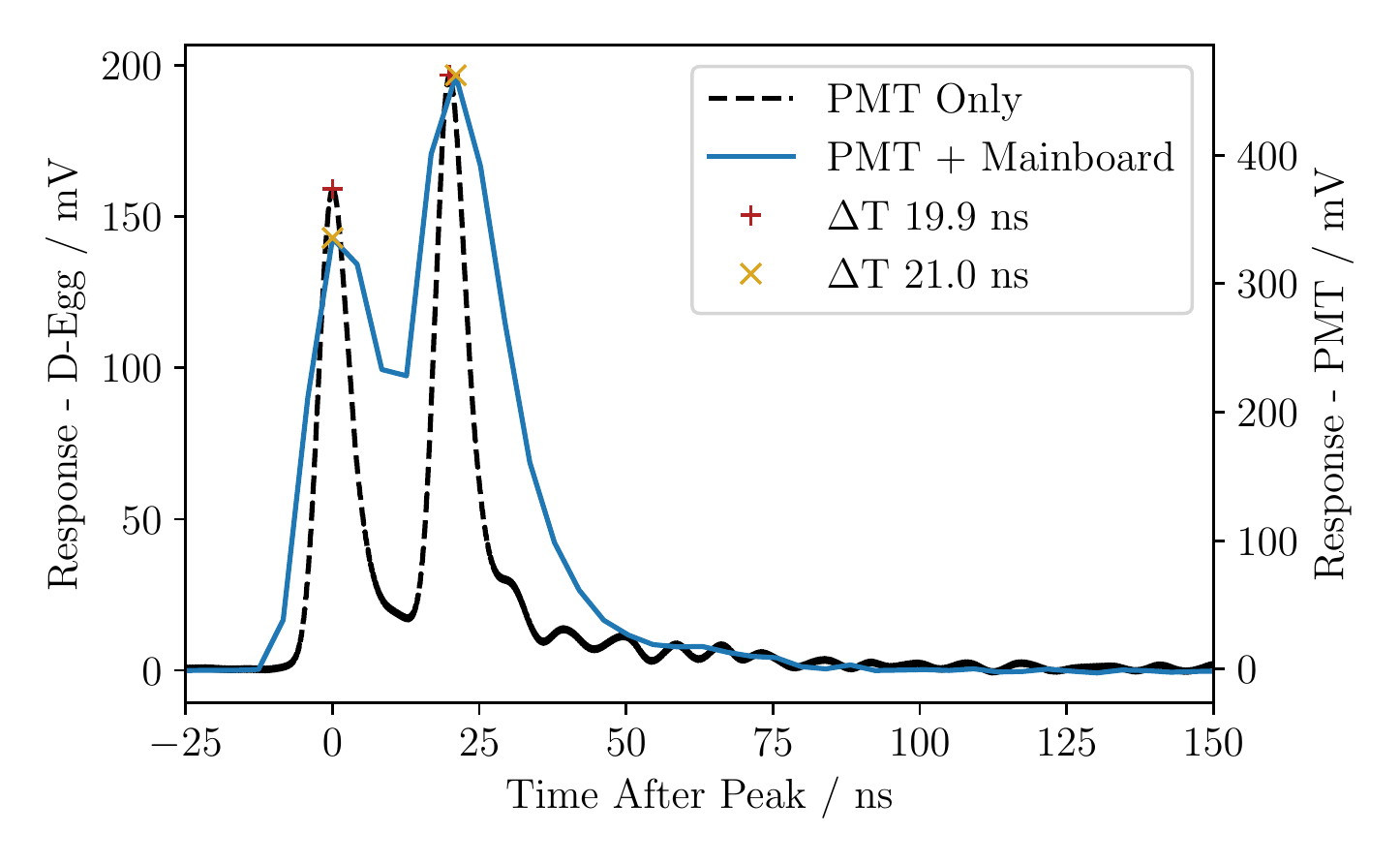}
\caption{Representative average of \num{5000} double-pulse waveforms for PMT-only (oscilloscope readout) and PMT-plus-mainboard configurations. Orange and red marks indicate the double-pulse peak structures identified in the waveforms and $\Delta$T the difference between the two peaks.}
\label{double_pulse_cold_temp}
\end{figure} 

Figure~\ref{double_pulse_cold_temp} is the average of \num{5000}~waveforms, including both PMT-only (oscilloscope readout) and PMT-plus-mainboard configurations. The key features are the two separated peaks, located using a simple peak-finding algorithm, indicated by orange and red crosses. Single-pulse triggers, primarily cosmic muons, were removed during the analysis stage. Based on the peak-finding algorithm, the separation between pulses is consistent with the timing accuracy of the mainboard. The two samples were aligned in time at the first identified peak to emphasize that although the electronics on the mainboard broaden the waveform shape, the peak positions are unchanged.


\section{Conclusions}\label{sec:conclusion}
The D-Egg is the first non-spherical optical module that has been proven to withstand the strict mechanical requirements for deployment in the South Pole ice as a part of the IceCube Upgrade. As a smaller module radius reduces the deployment cost, this design sets a precedent for future elongated optical modules to be deployed for the IceCube-Gen2 project and other experiments with stringent mechanical strength requirements. Optimal component selection contributes significantly to the overall D-Egg performance and cost-effectiveness. 
Improvements in the transparency of the pressure vessel glass and the optical silicone gel at short wavelengths allow significant improvements in detection efficiency.
Finally, compared with the single PMT IceCube DOM, the dual PMT design of the D-Egg improves the homogeneity of the sensitivity over the solid angle, which helps reduce the uncertainty in the photon-direction reconstruction. 
The D-Eggs have a Cherenkov spectrum averaged effective photodetection sensitivity that is \num{2.8}~times higher, with competitive dark noise rates and power consumption.

Completing the production of \num{310}~D-Eggs is a significant step forward for the IceCube Upgrade project, where mass testing prior to deployment allows careful characterization of individual modules as well as the overall performance of D-Eggs at low temperatures.
The in-lab module performance shows a stable operation at sustained temperatures equivalent to those at the South Pole. The performance during the temperature cycles ensures high survivability after deployment where drastic temperature transitions occur. In addition to operational stability, performance measurements confirmed that the D-Eggs meet the requirements for low-energy (\si{\giga\electronvolt}-scale) and high-energy (\si{\peta\electronvolt}-scale and up) physics. The mean dark rate of \SI{1.8}{\kilo\hertz} satisfies the projected dark rate triggering requirements that are most relevant for event reconstruction and low-energy analyses. Successful modeling of the D-Egg linearity, where the PMT and electronics contribute to the response, is critical for energy reconstruction, especially in high-energy studies. Furthermore, the reconstruction of event topologies and arrival directions relies heavily on identifying photon arrival times for which the typical D-Egg single-photon timing resolution of \SI{2.89}{\nano\second} is sufficient. The flexibility of the design, the simulated in-ice sensitivity, and the in-lab performance demonstrate that the D-Egg can enhance the sensitivity of the IceCube detector across multiple disparate physics areas.

\bibliographystyle{JHEP}
\bibliography{refs}

\acknowledgments
The IceCube collaboration acknowledges the significant contributions to this manuscript from Colton~Hill, Maximilan~Meier, Yasutsugu~Morii, Ryo~Nagai, and Nobuhiro~Shimizu.
The authors gratefully acknowledge the support from the following agencies:
USA {\textendash} U.S. National Science Foundation-Office of Polar Programs,
U.S. National Science Foundation-Physics Division,
U.S. National Science Foundation-EPSCoR,
Wisconsin Alumni Research Foundation,
Center for High Throughput Computing (CHTC) at the University of Wisconsin{\textendash}Madison,
Open Science Grid (OSG),
Extreme Science and Engineering Discovery Environment (XSEDE),
Frontera computing project at the Texas Advanced Computing Center,
U.S. Department of Energy-National Energy Research Scientific Computing Center,
Particle astrophysics research computing center at the University of Maryland,
Institute for Cyber-Enabled Research at Michigan State University,
and Astroparticle physics computational facility at Marquette University;
Belgium {\textendash} Funds for Scientific Research (FRS-FNRS and FWO),
FWO Odysseus and Big Science programmes,
and Belgian Federal Science Policy Office (Belspo);
Germany {\textendash} Bundesministerium f{\"u}r Bildung und Forschung (BMBF),
Deutsche Forschungsgemeinschaft (DFG),
Helmholtz Alliance for Astroparticle Physics (HAP),
Initiative and Networking Fund of the Helmholtz Association,
Deutsches Elektronen Synchrotron (DESY),
and High Performance Computing cluster of the RWTH Aachen;
Sweden {\textendash} Swedish Research Council,
Swedish Polar Research Secretariat,
Swedish National Infrastructure for Computing (SNIC),
and Knut and Alice Wallenberg Foundation;
Australia {\textendash} Australian Research Council;
Canada {\textendash} Natural Sciences and Engineering Research Council of Canada,
Calcul Qu{\'e}bec, Compute Ontario, Canada Foundation for Innovation, WestGrid, and Compute Canada;
Denmark {\textendash} Villum Fonden and Carlsberg Foundation;
New Zealand {\textendash} Marsden Fund;
Japan {\textendash} Japan Society for Promotion of Science (JSPS)
and Institute for Global Prominent Research (IGPR) of Chiba University;
Korea {\textendash} National Research Foundation of Korea (NRF);
Switzerland {\textendash} Swiss National Science Foundation (SNSF);
United Kingdom {\textendash} Department of Physics, University of Oxford.

\end{document}

%% file: authors.tex
\author[16]{R. Abbasi,}
\author[63]{M. Ackermann,}
\author[17]{J. Adams,}
\author[24]{N. Aggarwal,}
\author[11]{J. A. Aguilar,}
\author[21]{M. Ahlers,}
\author[22]{J.M. Alameddine,}
\author[30]{A. A. Alves Jr.,}
\author[44]{N. M. Amin,}
\author[42]{K. Andeen,}
% ...
\affiliation[11]{Universit{\'e} Libre de Bruxelles, Science Faculty CP230, B-1050 Brussels, Belgium}
% ...
\affiliation[16]{Department of Physics, Loyola University Chicago, Chicago, IL 60660, USA}
\affiliation[17]{Dept. of Physics and Astronomy, University of Canterbury, Private Bag 4800, Christchurch, New Zealand}
% ...
\affiliation[21]{Niels Bohr Institute, University of Copenhagen, DK-2100 Copenhagen, Denmark}
\affiliation[22]{Dept. of Physics, TU Dortmund University, D-44221 Dortmund, Germany}
% ...
\affiliation[24]{Dept. of Physics, University of Alberta, Edmonton, Alberta, Canada T6G 2E1}
% ...
\affiliation[30]{Karlsruhe Institute of Technology, Institute for Astroparticle Physics, D-76021 Karlsruhe, Germany }
% ...
\affiliation[42]{Department of Physics, Marquette University, Milwaukee, WI, 53201, USA}
% ...
\affiliation[44]{Bartol Research Institute and Dept. of Physics and Astronomy, University of Delaware, Newark, DE 19716, USA}
% ...
\affiliation[63]{Deutsches Elektronen-Synchrotron DESY, Platanenallee 6, 15738 Zeuthen, Germany }
% ...

%% file: extra_authors.tex
\affiliation[b]{Kobe Ocean-Bottom Exploration Center, Kobe University, Kobe 658-0022, Japan}
\affiliation[c]{also at Nippon Marine Enterprises, Yokosuka, Japan}